\documentclass[pdftex,twocolumn]{article}

\RequirePackage{graphicx}
\RequirePackage{mathptmx}      
\RequirePackage{flushend}
\RequirePackage[colorlinks,citecolor=blue,urlcolor=blue,linkcolor=blue]{hyperref}
\usepackage{cite}

\usepackage[utf8]{inputenc}
\usepackage[english]{babel}

\usepackage{amsmath}
\usepackage{amsfonts}
\usepackage{amssymb}
\usepackage{graphicx}
\usepackage{bm}
\usepackage{braket} 
\usepackage[usenames,dvipsnames,svgnames]{xcolor}
            
\usepackage{tikz}
\usepackage{pgfplots}
\usetikzlibrary{arrows,calc,external}

\usepackage{authblk}

\DeclareMathOperator{\Imag}{\mathrm{Im}}
\DeclareMathOperator{\Real}{\mathrm{Re}}
\DeclareMathOperator{\Tr}{\mathrm{Tr}}

\newcommand{\mA}{\mathcal{A}}

\newcommand{\mD}{\mathcal{D}}
\newcommand{\mF}{\mathcal{F}}
\newcommand{\mG}{\mathcal{G}}

\newcommand{\mL}{\mathcal{L}}

\newcommand{\mN}{\mathcal{N}}
\newcommand{\mO}{\mathcal{O}}
\newcommand{\mQ}{\mathcal{Q}}

\newcommand{\mU}{\mathcal{U}}

\begin{document}

\title{Top-antitop production from $W^+_L W^-_L$ and $Z_L Z_L$ scattering \\
under a strongly-interacting symmetry-breaking sector}


\author[1]{Andr\'es Castillo\thanks{afcastillor@unal.edu.co}}
\author[2]{Rafael L. Delgado\thanks{rdelgadol@ucm.es}}
\author[3]{Antonio Dobado\thanks{fllanes@fis.ucm.es}}
\author[3]{Felipe J. Llanes-Estrada\thanks{dobado@fis.ucm.es}}

\affil[1]{%
Universidad Nacional de Colombia, Sede Bogot\'a, Facultad de Ciencias, Departamento de F\'{\i}sica.\\
Ciudad Universitaria 111321, Bogot\'a, Colombia}
\affil[2]{%
Departamento de F\'{\i}sica Te\'orica I,
Universidad Complutense de Madrid, E-28040 Madrid, Spain;
on leave at SLAC, 2575 Sand Hill Rd, Menlo Park, CA 94025, USA}
\affil[3]{%
Departamento de F\'{\i}sica Te\'orica I,
Universidad Complutense de Madrid, E-28040 Madrid, Spain}


\maketitle 

\begin{abstract}
By considering a Non-linear Electroweak Chiral Lagrangian, including the Higgs, coupled to heavy quarks, and the Equivalence Theorem, we compute the one-loop scattering amplitudes $W^+W^-\to t\bar t$, $ZZ\to t\bar t$ and $hh\to t\bar t$ (in the regime $M_t^2/v^2\ll\sqrt{s}M_t/v^2\ll s/v^2$ and to NLO in the effective theory). 
We calculate the scalar partial-wave helicity amplitudes which allow us to check unitarity at the perturbative level in both $M_t/v$ and $s/v$. As with growing energy  perturbative unitarity deteriorates,  we also introduce a new unitarization method with the right analytical behavior on the complex $s$-plane and that can support poles on the second Riemann sheet to describe resonances in terms of the Lagrangian couplings. Thus we have achieved a consistent phenomenological description of any resonant $t\bar{t}$ production that may be enhanced by a possible strongly interacting Electroweak Symmetry Breaking Sector.
\end{abstract}
\section{Introduction}
The Higgs-like particle with a mass of $125\,{\rm GeV}$ found at the Large Hadron Collider (LHC)~\cite{ATLAS,CMS} completes a possible framework of the fundamental interactions, as this new boson has quantum numbers and couplings compatible with those expected for the Higgs of the Standard Model (SM) in its minimal version. In addition, new scalar-resonances associated to new physics effects have been constrained roughly up to 600-700 GeV~\cite{twophotons}. 
For new vector bosons, the lowest energy for a possible resonance to lie at is even higher~\cite{searches}. The discrepancy among the Higgs mass scale and that of any new physics appearance is suggestive of a Goldstone boson (GB) interpretation of the Higgs that (together with the Goldstone bosons associated with the $W^{\pm}_L$ and $Z_L$ components of vector bosons), may be related to some global spontaneous symmetry breaking that in turn prompts a breaking of the electroweak gauge symmetry $SU(2)_L\times U(1)_Y\to U(1)_Q$.

To describe such pseudo-Goldstone behavior of the Higgs boson, some effective description of the Electroweak Symmetry Breaking Sector (EWSBS) of the SM must be taken into account \cite{Espriu:2013fia,Brivio:2013pma,Azatov:2012bz,Alonso:2012px,Pich:2013fba,Jenkins:2013zja,Degrande:2012wf,Buchalla:2013rka,Buchalla:2012qq}. These Effective Field Theory (EFT) descriptions are useful even when the Higgs is not a GB. In consequence, EFTs are a convenient way of parametrizing the EWSBS.

The energy gap may also favor a non-linear Lagrangian description of the symmetry breaking, which is a very general approach to the EWSBS in the EFT. The old Electroweak Chiral Lagrangian (ECL) technique~\cite{Appelquist}, built up on standard Chiral Perturbation Theory for hadron physics~\cite{ChPT}, can be extended to include the scalar Higgs-like particle $h$ transforming as a singlet of custodial $SU(2)_C$ to give the so-called Higgs Effective Field Theory (HEFT). Meanwhile, the longitudinal gauge bosons transform as a triplet. This pattern is analogous to low energy hadron physics, where pions fall in a triplet and the $\eta$ meson is embedded in a singlet representation of the strong $SU(2)_V$ isospin group. The global symmetry breaking scheme, $SU(2)_L\times SU(2)_R\to SU(2)_C$, is common to both effective field theories of the strong and electroweak interactions.

As the HEFT theories are derivative expansions, for most of parameter space (saliently excluding that of the Standard Model and perhaps other very carefully tuned sets), the interactions will generically become strong at sufficiently high energy, and we have argued that a second, very broad scalar pole is expected \cite{Delgado:2013hxa, Delgado:2014dxa}. This motivates theoretical studies of new resonances with energies $700\,{\rm GeV}<E<4\pi v\sim 3\,{\rm TeV}$ that require methods extending perturbation theory in the HEFT Lagrangian --that we explote  to Next to Leading Order, (NLO)--. One strategy is extending the low-energy amplitudes through   dispersion relations (DR) compatible with analyticity and unitarity. Resonances can then be found as poles in the second Riemann sheet due to the proper analytical behavior of the amplitudes.

Such unitarization methods introduce some level of arbitrariness, as unitarity, analyticity, and the low-energy behavior are not sufficient to determine a scattering amplitude with arbitrary accuracy. Nevertheless, in \cite{Delgado:2015kxa} we showed that the analytical and unitary description of higher energy dynamics provided by DRs extending the one-loop results, is essentially unique qualitatively; at least so up to the first resonance in each spin-isospin channel. Other groups have recently pursued related unitarization methods in the context of the EWSBS~\cite{unitarization}.

The top quark is quite strongly coupled to the EWSBS and offers an opportunity for numerous analysis~\cite{Franceschini:2017eyj,Djouadi:2016ack,Hespel:2016qaf,Czakon:2017wor}. Current experimental efforts have studied in detail processes where heavy quarks are produced as intermediate (subsequently decaying into jets) or final states \cite{Aad:2016zqi,Fabbri:2017irk,Khachatryan:2016gxp,Jeon:2017fas}.  It is then reasonable to introduce fermions in the theory for energy scales compatible with those where  resonances may appear at the LHC and may be described within the EFT framework. 
Our work analyzes the coupling of the pure Goldstone sector to top quarks. New-physics fermionic couplings in the HEFT entitle us to flexibly describe the amplitudes $ W^+_L W^-_L \to t\bar{t}$, $Z_L Z_L\to t\bar{t}$ and $hh\to t\bar{t}$ in the regime $M_t^2/v^2\ll \sqrt{s}M_t/v^2\ll s/v^2$.  
In the high energy limit $s\gg M_{Z,W}^2\sim M_h^2$ and by means of the Equivalence Theorem (ET) \cite{ET}, we can compute all amplitudes $V_L V_L\to t\bar{t}$ substituting the $V_L$ longitudinal vector bosons by GBs (denoted $\omega$ in what follows). 
Therefore, in this article we will take $0=M_W^2=M_Z^2=M_h^2$ consistently. 
The last equality, for $M_h$, can also be a consequence of a new symmetry breaking pattern (such as in Composite Higgs Models) and is within the philosophy of EFT, but holds (approximately) anyway because of the experimental Higgs mass value, which is close to that of the electroweak gauge bosons, below the TeV scale that we explore. 

Watson's final state interaction theorem, implemented in our unitarization method, guarantees that amplitudes with final $t\bar{t}$ pairs feature poles in the second Riemann sheet in the same position as the elastic GB amplitudes. Dynamical resonances are thus linked to the parameter space of chiral couplings in the DR-unitarized HEFT.

We have organized the presentation as follows: section \ref{sec:EWChTFermions} discusses the introduction of a heavy fermion in the Effective Lagrangian for Electroweak-chiral interactions. The amplitudes for $\omega^i\omega^j\to t\bar{t}$ and $hh\to t\bar{t}$ processes at tree and one-loop levels are computed and summarized in sections~\ref{sec:amplitudeswTLOl} and~\ref{sec:amplitudeshTLOl}, respectively. We dedicate sections~\ref{sec:helicity} and \ref{sec:partialwaves} to the helicity amplitudes and to the scalar partial-wave computation. 
Section~\ref{sec:unitarization} is dedicated to the study of unitarity and the Inverse Amplitude Method  (IAM~\cite{Dobado:1989gr,Dobado:1989qm,Dobado:1992ha}) implementation, both for single and coupled channels (while a derivation of the later is deferred to an appendix). Section~\ref{sec:discussion} offers our final remarks and discussion.

\section{The Electroweak Chiral Lagrangian with massive fermions}
\label{sec:EWChTFermions}
There are several equivalent forms of the universal Electroweak Chiral Lagrangian employing only the experimentally known particles. At leading order we adopt the gauged $SU(2)_L\times SU(2)_R/SU(2)_C=SU(2)\simeq S^3$ HEFT. Adding the chiral fermionic interactions, the Lagrangian reads
\begin{multline}  \label{LagrangianI}
\mL = \frac{v^2}{4}\mF(h)\Tr\left[\left(D_\mu U\right)^\dagger D^\mu U\right] %
      +\frac{1}{2}\partial_\mu h \partial^\mu h   \\
      -  V(h) + i\bar{Q}\partial Q 
      - v\mG(h)\left[\bar{Q}_L^\prime UH_Q Q_R^\prime + \text{h.c.} \right],
\end{multline}
where the $U(x)\in SU(2)$ can be parametrized in terms of the would-be Goldstone fields as
\begin{equation}
U=\sqrt{1-\frac{\omega^2}{v^2}}+i\frac{\bar{\omega}}{v},
\end{equation}
with $\bar{\omega}=\tau_i\omega^i$.
The $SU(2)_L\times U(1)_Y$ covariant derivative is given by
\begin{equation}
D_\mu U = \partial U_\mu-ig\frac{\sigma_i}{2} W_\mu^i U + ig^\prime U\frac{\sigma_3}{2} B_\mu .
\end{equation}
The Higgs potential can be expanded as
\begin{equation} \label{Higgspotential}
  V(h)\, = \, v^4\;\sum_{n=3}^\infty V_n \left(\frac{h}{v}\right)^n\ .
\end{equation}
We recover the SM with $V_3 = \frac{M_h^2}{2v^2}$, $V_4 = \frac{M_h^2}{8v^2}$, 
$V_{n>4} = 0$.
In most models of interest that the low energy theory formulated as HEFT is supposed to describe,
the coefficients of the Higgs self-potential scale in the same way, as powers of the Higgs mass.
It is a reasonable hypothesis to maintain this scaling as the constraints on these couplings 
have so far been found to be close to their SM values.
As discussed in the introduction, we are neglecting $M_h$~\cite{Delgado:2013loa} because we work in the $M_W^2\sim M_h^2 \ll s$ limit, 
therefore the potential $V(h)$ is negligible and we further set it to zero. 

In the Yukawa sector (last line) of the Lagrangian in Eq.~(\ref{LagrangianI}), the quark doublets are%
\begin{equation}
Q^{(\prime)} = \left( %
    \begin{array}{c}
        \mU^{(\prime)} \\ 
        \mD^{(\prime)}
    \end{array} %
\right),
\end{equation}
where the two $Q$ entries are made of the different up and down quark sectors%
\begin{align}
\mU^\prime &=\left( u,c,t\right)^\prime,  &
\mD^\prime &=\left( d,s,b\right)^\prime,
\end{align}
and the Yukawa-coupling matrix in Eq.~(\ref{LagrangianI}) has the form%
\begin{equation}
H_Q=\left( 
      \begin{array}{cc}
        H_U & 0 \\ 
        0   & H_D%
      \end{array}%
\right) .
\end{equation}
This matrix can be diagonalized by transforming independently the right- and left-handed up and down quarks as
\begin{align}
\mD_{L,R} = V_{L,R}^D\mD_{L,R}^\prime , && \mU_{L,R} = V_{L,R}^U \mU_{L,R}^\prime ,
\end{align}
where $V_{L,R}^{U,D}$ are four $3\times 3$ unitary matrices. Thus the Yukawa part of the Lagrangian can be written as
\begin{align}
\mL_Y ={}&-\mG\left( h\right) \left\{%
   \sqrt{1-\frac{\omega^2}{v^2}} \left( \overline{\mU}M_U\mU + \overline{\mD}M_D\mD\right)              \right. \nonumber\\  
  &\left. +\frac{i\omega^0}{v}\left( \overline{\mU}M_U\gamma^5\mU - \overline{\mD}M_D\gamma^5\mD\right) \right. \nonumber\\
  &\left. +i\sqrt{2}\frac{\omega^+}{v}\left(\overline{\mU}_L V_{CKM} M_D\mD_R-\overline{\mU}_R M_U V_{CKM}\mD_L\right)  \right. \nonumber\\
  &\left. +i\sqrt{2}\frac{\omega^-}{v}\left(\overline{\mD}_L V_{CKM}^\dagger M_U\mU_R-\overline{\mD}_R M_DV_{CKM}^\dagger \mU_L\right)  \right\} 
\end{align}
where $\omega^\pm =(\omega^1 \mp i \omega^2)/\sqrt{2}$,  $\omega^0 = \omega^3$, $V_{CKM}= V_{L,}^U V_{L,}^{D\dagger}$ is the Cabibbo-Kobayashi-Maskawa matrix  and the new quark fields are mass eigenstates with $M_{U}$ and $M_{D}$ being the corresponding diagonal and real mass matrices. 

In keeping with the $m_t/v\ll\sqrt{s}/v$ philosophy, lighter quark masses are completely irrelevant so we focus  only on the heaviest quark generation, for which
\begin{align}
\mL_Y ={}&-\mG\left( h\right) \left\{%
    \sqrt{1-\frac{\omega^2}{v^2}}\left( M_t t\bar{t} + M_b\bar{b}b \right)              \right. \nonumber\\
  &\left.  +\frac{i\omega^0}{v}\left(M_t \bar{t}\gamma^5 t -M_b\bar{b}\gamma^5 b\right) \right. \nonumber\\
  &\left. +i\sqrt{2}\frac{\omega^+}{v}\left( M_b\bar{t}_L b_R - M_t\bar{t}_R b_L\right)  \right. \nonumber\\
  &\left. +i\sqrt{2}\frac{\omega^-}{v}\left( M_t\bar{b}_L t_R - M_b\bar{b}_R t_L\right)  \right\} .  \label{LY1G}
\end{align}
where the matrix element  $V_{tb}$ has now been omitted since it is very close to unity. 
As can be seen, this part of the Lagrangian explicitly breaks custodial symmetry because of the (very) different values of the $t$ and $b$ quark masses. 

The $\mF$ and $\mG$ functions appearing in the Lagrangian are arbitrary analytical functions on the Higgs field $h$, which are usually parametrized as
\begin{equation}
     \mF\left( h\right) = 1 + 2a\frac{h}{v} + b\frac{h^2}{v^2}+\dots
\end{equation}
and
\begin{equation}
     \mG\left( h\right) = 1 + c_1\frac{h}{v} + c_2\frac{h^2}{v^2}+\dots
\end{equation}
In this work, these functions are only needed up to the quadratic terms. Also we will consider the limit of vanishing mass for the bottom quark ($M_b=0$). Then, the expanded Yukawa Lagrangian is
\begin{align}
\mL_Y ={}& -\left( 1+c_1\frac{h}{v}+c_2\frac{h^2}{v^2}\right) \left\{%
     \left( 1-\frac{\omega^2}{2v^2}\right) M_t t\bar{t}\right. \nonumber\\
    &\left. +\frac{i\omega^0}{v}M_t\bar{t}\gamma^5 t - i\sqrt{2}\frac{\omega^+}{v} M_t\bar{t}_R b_L%
        +i\sqrt{2}\frac{\omega^-}{v} M_t\bar{b}_L t_R\right\} ,
\end{align}
where we have kept only $\mO(\omega^2)$ terms. 

Finally, the relevant HEFT Lagrangian that couples the EWSBS to the 3rd fermion generation, so as to describe the $\omega\omega\to t\bar{t}$ and $hh\to t\bar{t}$ processes in the regime $M_t^2/v^2\ll M_t\sqrt{s}/v^2\ll s/v^2$, is given by 
\begin{align} \label{EFTLagrangianexpanded}
\mL &=\frac{1}{2}\partial_\mu h\partial^\mu h -\left(1+c_1\frac{h}{v}+c_2\frac{h^2}{v^2}\right)
\left\{ \left(1-\frac{\omega^2}{2v^2}\right) M_t t\bar{t} \right.\notag\\
    &\left.+ \frac{i\sqrt{2}\omega^0}{v} M_t\bar{t}\gamma^5 t-i\sqrt{2}\frac{\omega^+}{v}M_t\bar{t}_R b_L+i\sqrt{2}\frac{\omega^-}{v}M_t \bar{b}_L t_R\right\}  \notag\\
&+\frac{1}{2}\left(1+2a\frac{h}{v}+b\left(\frac{h}{v}\right)^2\right)
\partial_\mu\omega^i\partial^\mu\omega_j\left(\delta_{ij}+\frac{\omega_i\omega_j}{v^2}\right) .
\end{align}
As we will see later, in order to properly unitarize the $\omega\omega \to t\bar{t}$ and $hh\to t\bar{t}$ amplitudes, one has to consider also the amplitudes $\omega\omega \to \omega\omega$, $\omega \omega \to hh$ and $hh\to hh$. The one-loop divergences appearing  in all of them can be absorbed in the couplings corresponding to the Lagrangian (as we will explicitly show for the top amplitudes later)
\begin{align}
\mL_4 ={}&%
    \frac{4a_4}{v^4}\partial_\mu\omega^i\partial_\nu\omega^i\partial^\mu\omega^j\partial^\nu\omega^j%
   +\frac{4a_5}{v^4}\partial_\mu\omega^i\partial^\mu\omega^i\partial_\nu\omega^j\partial^\nu\omega^j%
\nonumber\\
&  +\frac{2d}{v^4}\partial_\mu h\partial^\mu h\partial_\nu\omega^i\partial^\nu\omega^i%
   +\frac{2e}{v^4}\partial_\mu h\partial^\nu h\partial^\mu\omega^i\partial_\nu\omega^i%
\nonumber\\
&  + \frac{g}{v^4}\left(\partial_\mu h\partial^\mu h\right)^2%
\nonumber\\
&  +g_t\frac{M_t}{v^4}(\partial_\mu\omega^i\partial^\mu\omega^j) t\bar{t} %
   +g'_t \frac{M_t}{v^4}(\partial_\mu h\partial^\mu h) t\bar{t}.
\end{align}

\section{Tree level and one loop contributions for $\protect\omega^{a}\protect\omega^{b}\to t\bar{t}$} \label{sec:amplitudeswTLOl}

\begin{figure}
\null\hfill\includegraphics[width=.3\textwidth]{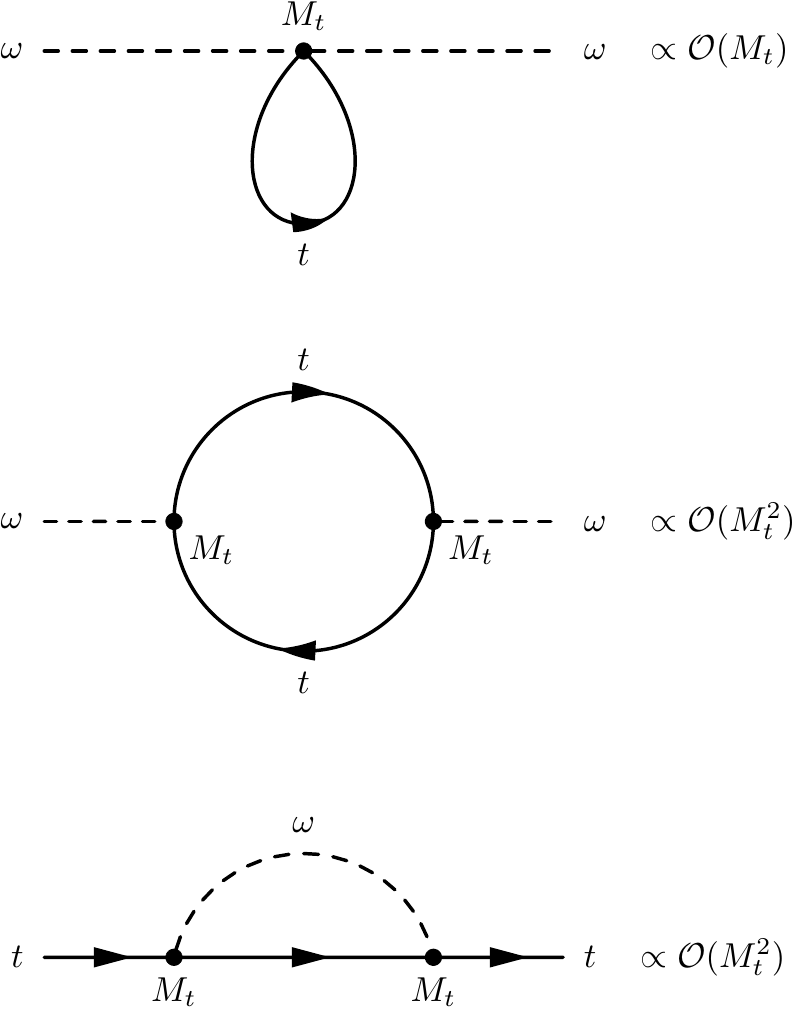}\hfill\null
\newline
\caption{\label{fig:diagramassuprimidos} The contribution of these diagrams to $\omega\omega\to t\bar{t}$ scattering (through wavefunction or mass renormalization) does not need to be considered, as all our amplitudes are already linear in $M_t$. Attaching any of these corrections would increase the order in the $M_t/v$ expansion by at least one unit, as exposed in Fig.~\ref{fig:counting}. }
\end{figure}

In this section we address the process $V_LV_L \to t\bar{t}$ (where $V=W,Z$) at energies that are high when compared with $M_Z$, $M_W$ and $M_h$; then we can use the ET and concentrate only in the GB $\omega^i$, $h$ and the $b$ and $t$ quarks. More specifically, we will consider the regime $M_t^2/v^2\ll\sqrt{s}M_t/v^2\ll s/v^2$. 
In earlier work we have set all masses to zero from the start, $M_h=M_Z=M_W=M_t=0$ since we were interested in the high energy regime, appropriate for LHC resonance searches. However in this work we deal with $t\bar{t}$ production and in that strict limit the amplitude vanishes and the minimal non-vanishing contribution must be at least linear in $M_t$. 
More precisely, the lowest order (tree level) $V_LV_L \to t\bar{t}$ is of the order of $\sqrt{s}M_t/v^2$. At the one-loop level one should in principle include diagrams with $\omega$, $h$ and $t$ loops. However, diagrams with $t$ loops are higher order in $M_t/v$. Thus, if one is interested in the region $M_t^2/v^2\ll\sqrt{s}M_t/v^2$, diagrams with $t$ loops can safely be ignored, even if $M_t/v$ is not a very small parameter. On the other hand, one-loop diagrams with $\omega$ and $h$ loops are order $\sqrt{s}M_t/v^2$, as the tree level ones, and must consistently be taken into account. Consequently, we will ignore diagrams as those in Fig.~\ref{fig:diagramassuprimidos}. This will not only make the computation manageable (because of the significantly smaller number of Feynman diagrams to be taken into account) but also it will make the renormalization of the amplitudes much simpler, so that only two new counter-terms must be introduced and the corresponding two couplings renormalized. We consider this a very sensible approximation to the $V_LV_L \to t\bar{t}$ reaction in the $M_t^2/v^2\ll\sqrt{s}M_t/v^2\ll s/v^2$ regime and, in any case, a necessary first step to a more complete future computation that should be performed if more accuracy was ever needed.

\begin{figure}[h]
\centering
\includegraphics[width=.9\columnwidth]{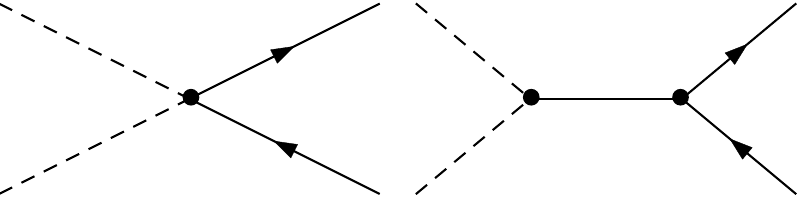}\newline
\caption{Non-vanishing tree level contributions to the process $\protect\omega\protect\omega\to t\bar{t}$. Dashed lines represent the $\omega$ Goldstone bosons. The continuous line is the Higgs. The arrowed, continuous line stand for $t$ and $\bar{t}$. }
\label{fig:Tree}
\end{figure}

 At tree level (see Fig.~\ref{fig:Tree}), the scattering amplitude is given by
\begin{multline}
\mQ^{\text{tree}} \left(\omega^i\omega^j\to t^{\lambda_1}\bar{t}^{\lambda_2}\right)  \\
 = \sqrt{3}\left(1-ac_1+\frac{g_t}{2}\frac{s}{v^2}\right) \frac{M_t}{v^2} \bar{u}^{\lambda_1}(p_1)v^{\lambda_2}(p_2)\delta^{ij}, \label{GoldGold1}
\end{multline}
where $i$, $j$ are the custodial isospin indices of the incoming GB; $p_1$, $p_2$ and $\lambda_1$, $\lambda_2$ are top, antitop momenta and helicities respectively. The $\sqrt{3}$  factor is a color factor since the $t\bar t$ pair is produced in a color singlet state. 

\begin{figure*}
\null\hfill\includegraphics[width=.6\textwidth]{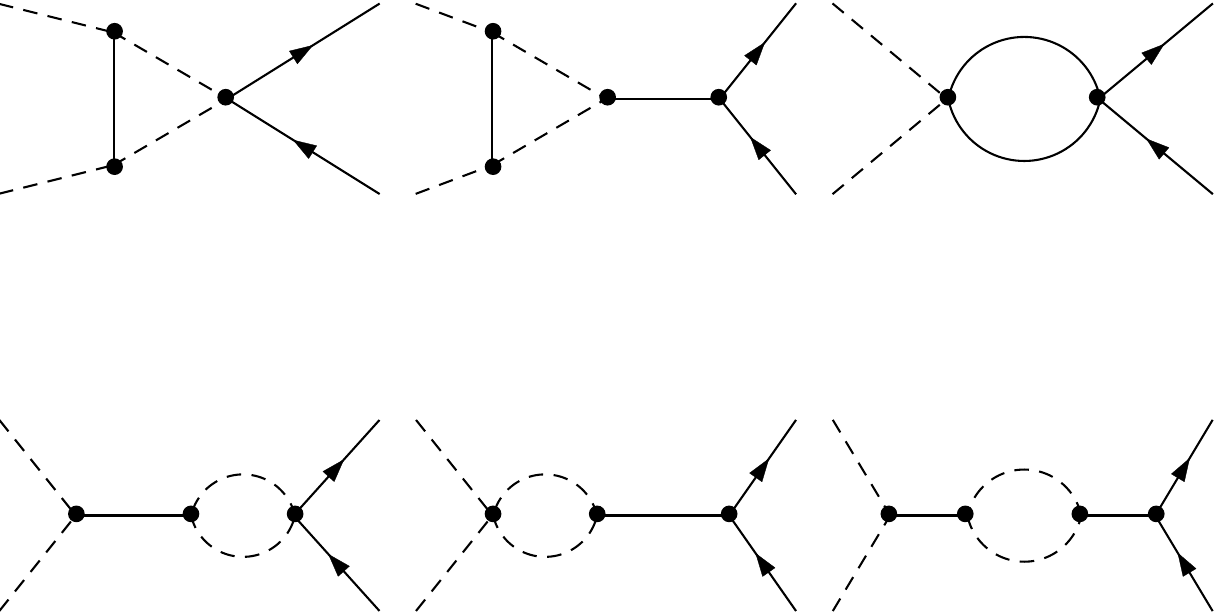}\hfill\null\newline
\caption{One loop contributions for $\protect\omega \protect\omega \to t\bar{t}$.}
\label{fig:OneWW}
\end{figure*}

Next we consider the one-loop terms. The Feynman diagrams contributing to this order can be seen in Fig.~\ref{fig:OneWW}. By using dimensional regularization with dimension $D=4-\epsilon$, the result is
\begin{multline}
\mQ^{\text{1-loop}}  \left( \omega^i\omega^j\to t^{\lambda_1}\bar{t}^{\lambda_2}\right)  \\
= \sqrt{3}\frac{s}{(4\pi)^2 v^2}\left[\left( 1-ac_1\right) \left( 1-a^2\right) + c_2\left( b-a^2\right) \right] \\   
\times \left( N_\varepsilon+2-\log\frac{-s}{\mu^2}\right)\frac{M_t}{v^2}\bar{u}^{\lambda_1}v^{\lambda_2}\delta^{ij},
\end{multline}
where as usual
\begin{equation}
     N_{\varepsilon}=\frac{2}{\varepsilon}+\log 4\pi - \gamma,
\end{equation}
and $\mu$ is an arbitrary renormalization scale. Hence, the sum of the two contributions is%
\begin{equation}
  \mQ\left(\omega^i\omega^j\to t^{\lambda_1}\bar{t}^{\lambda_2}\right) = %
      \sqrt{3}\left(Q^{\text{tree}}+Q^{\text{1-loop}}\right)\frac{M_t}{v^2}\bar{u}^{\lambda_1}v^{\lambda_2}\delta^{ij},
\end{equation}
with
\begin{align}
Q^{\text{tree}}(s)   &= 1-ac_1 + \frac{g_t}{2}\frac{s}{v^2}  \label{res_Q_tree}\\
Q^{\text{1-loop}}(s) &= \frac{s}{(4\pi)^2 v^2}C_t \left(N_{\varepsilon}+2-\log\frac{-s}{\mu^2}\right) \label{res_Q_NLO} 
\end{align}
and%
\begin{equation}
C_t = (1-ac_1)(1-a^2)+c_2 (b-a^2) .
\end{equation}
The divergence in Eq.~(\ref{res_Q_NLO}) can be absorbed by renormalizing the $g_t$ coupling.  Using  the  $\overline {MS}$ renormalization scheme we define
\begin{equation}
   g_t^r=g_t+\frac{C_t}{8\pi^2}N_\epsilon
\end{equation}
and consequently the next to leading order contribution (NLO) is given by
\begin{align} \label{Qdecomposed}
Q^{\rm NLO}(s) &= Q^{\text{tree}}\left(s\right)+Q^{\text{1-loop}}\left(s\right) \\  \nonumber 
               &= 1-ac_{1}+\frac{s}{v^2}\left[\frac{g_t^r}{2}+\frac{C_t}{(4\pi)^2}\left(2-\log\frac{-s}{\mu^2}\right)\right] .
\end{align}
Notice that we do not have any wavefunction or mass renormalization. To see it, observe the typical diagrams contributing in Fig.~\ref{fig:diagramassuprimidos}. 
All diagrams in the figure are clearly proportional to $M_t$ or higher orders. Since all our amplitudes have been computed to linear order in $M_t\sqrt{s}/v^2 \ \ (\ll s/v^2)$, attaching them to any of our external legs or propagators would increase the order in the chiral $M_t\sqrt{s}/v^2$ counting of Fig.~\ref{fig:counting}.

In the absence of wave or mass renormalization, squared amplitudes must be observable and hence $\mu$-independent. Then, we require the total derivatives of the NLO amplitude with respect to $\log\mu^2$ to vanish,
\begin{equation}
\frac{dQ^{\rm NLO}(s)}{d\log\mu^2} = \frac{s}{v^2}\left[\frac{1}{2}\frac{d g_t^r}{d\log\mu^2}+\frac{C_t}{(4\pi)^2}\right] = 0.
\end{equation}
Hence, the renormalization equation for the coupling $g_t^r$ reads
\begin{equation}
\frac{d g^r_t}{d\log\mu^2}=-\frac{C_t}{8\pi^2},
\end{equation}
which can be integrated to give
\begin{equation}
g^r_t\left(\mu\right) = g^r_t\left(\mu_0\right)-\frac{C_t}{8\pi^2}\log\left(\frac{\mu^2}{\mu_0^2}\right) .
\end{equation}
On the other hand, the different spinor combinations of helicities appearing in the above amplitudes are, to the Leading Order (LO) in $M_t/\sqrt{s}$ expansion,
\begin{align}
\bar{u}^+ (p_1) v^{+}(p_2) &=  \sqrt{s-4M_t^2}\simeq\sqrt{s}  \nonumber \\
\bar{u}^+ (p_1) v^{-}(p_1) &=  0  \nonumber \\
\bar{u}^- (p_1) v^{+}(p_2) &=  0  \nonumber \\
\bar{u}^- (p_1) v^{-}(p_2) &= -\sqrt{s-4M_t^2}\simeq -\sqrt{s},
\end{align}%
where the helicity indices $+$ and $-$ refer to $\lambda = +1/2$ and $\lambda = - 1/2$, respectively. Therefore, the tree level amplitude in Eq.~(\ref{res_Q_tree}) is of order $\mO(\sqrt{s}M_t/v^2)$ and the one-loop in Eq.~(\ref{res_Q_NLO}), of order $\mO(s\sqrt{s}M_t/v^4)$.

Thus, the amplitude $\omega^i\omega^j\to t\bar t$ is given by
\begin{align}
\mQ\left(\omega^i\omega^j\to t^+\bar{t}^+\right)
  &=\sqrt{3}Q^{\rm NLO}(s)\frac{M_t\sqrt{s}}{v^2}\delta^{ij} \label{AmplitudeGoldstone}\\
\mQ\left(\omega^i\omega^j\to t^-\bar{t}^-\right)
  &=-\mQ\left(\omega^i\omega^j\to t^+\bar{t}^+\right) \label{AmplitudeGoldstone2} \\
\mQ\left(\omega^i\omega^j\to t^-\bar{t}^+\right)
  &= \mQ\left(\omega^i\omega^j\to t^+\bar{t}^-\right) = 0. \label{AmplitudeGoldstone3}
\end{align}

\section{$hh\to t\bar{t}$ process}
\label{sec:amplitudeshTLOl}

\begin{figure}[tph]
\centering
\includegraphics[scale=0.8]{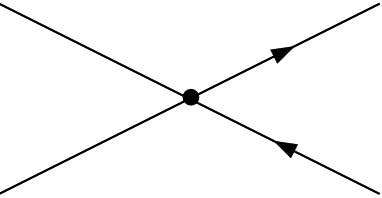} \vspace{0.75cm}
\caption{Contribution at LO to $hh\to t\bar{t}$ annihilation}
\label{fig:Higgs} 
\end{figure}

In a similar way to the triplet-states annihilation, we may consider $hh\to t\bar{t}$ annihilation. 
The contributing diagram (direct vertex) to the LO $hh\to t\bar{t}$ amplitude is depicted in Fig.~\ref{fig:Higgs} and the tree level amplitude is easily obtained,
\begin{multline}
 \mN^{\rm tree} \left( hh\to t^{\lambda_1}\bar{t}^{\lambda_2}\right)=    \\ 
  \sqrt{3}\left(-\frac{2c_2 M_t}{v^2} +\frac{g'_t s M_t}{2v^4} \right)\bar{u}^{\lambda_1}\left(p_1\right) v^{\lambda_2}\left(p_2\right)     
\end{multline}

\begin{figure*}
\centering
\includegraphics[width=.9\textwidth]{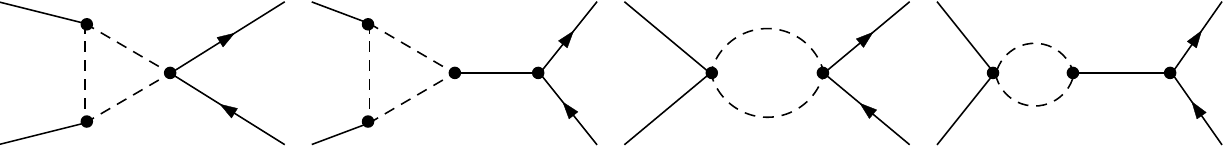}\newline
\caption{One-loop contributions to $hh\to t\bar{t}$ amplitude.}
\label{fig:Higgoneloop2}
\end{figure*}

At the one-loop level the amplitude is given by the diagrams in Fig.~\ref{fig:Higgoneloop2} which add up to
\begin{multline}
\mN^{\rm 1-loop}\left(hh\to t^{\lambda_1}\bar{t}^{\lambda_2}\right) %
 =  -\sqrt{3}\frac{3 s M_t}{32\pi^2 v^4}(b-a^2)(1-ac_1)  \\ 
   \times \left( N_{\varepsilon }+2-\log \frac{-s}{%
\mu ^{2}}\right) \bar{u}^{\lambda_1}\left( p_{1}\right) v^{\lambda_2}\left( p_{2} \right). \label{Nza}
\end{multline}

Combining the tree plus the one-loop amplitude yields
\begin{multline}
\mN^{\rm NLO} \left( hh\to t^{\lambda_1} \bar{t}^{\lambda_2}\right) = \\
 \sqrt{3}\left\{%
     -2c_2 + \frac{s}{v^2}\left[\frac{g'_t }{2} - \frac{3}{32\pi^2} C_t'\left(N_\varepsilon +2-\log \frac{-s}{\mu^2}\right)\right]
 \right\} \\
  \times\bar{u}^{\lambda_1}\left(p_1\right) v^{\lambda_2}\left(p_2\right)  
\end{multline}
or
\begin{multline}
\mN^{\rm NLO} \left( hh\to t^{\lambda_1} \bar{t}^{\lambda_2}\right) = \\
\sqrt{3}\left\{ -2c_2 +\frac{s}{v^2}\left[\frac{g'^r_t}{2} - \frac{3}{32\pi^2} C_t'
 \left(2-\log\frac{-s}{\mu^2}\right)\right]\right\}  \\ 
 \times \bar{u}^{\lambda_1}\left(p_1\right) v^{\lambda_2}\left(p_2\right),
\label{AmplitudeII}
\end{multline}
where the renormalized coupling $g'^r_t$ is obviously defined as
\begin{equation}
g'^r_t = g'_t-\frac{3C'_t}{(4\pi)^2}N_\varepsilon
\end{equation}
and
\begin{equation}
C'_t=(b-a^2)(1-ac_1).
\end{equation}

Again, the lack of wave function renormalization at this level requires this amplitude to be scale independent. Thus the coupling dependence on $\mu$ is given by
\begin{equation}
\frac{d g'^r_t}{d\log\mu^2}=\frac{3C'_t}{(4\pi)^2}.
\label{BetaR}
\end{equation}

By integrating Eq.~(\ref{BetaR}), the renormalized coupling evolves with the scale as%
\begin{equation}
g'^r_t\left(\mu\right) = g'^r_t(\mu_0)+\frac{3C'_\text{t}}{(4\pi)^2}\log\left(\frac{\mu^2}{\mu_0^2}\right) .
\end{equation}

In a similar way than in the would-be GB case we get, for the amplitude in Eq.~(\ref{AmplitudeII}),%
\begin{align}
 \mN\left( hh\to t^-\bar{t}^-\right) &= -\mN\left(hh\to t^+\bar{t}^+\right) , \label{AmplitudeHiggsIIpre}\\
 \mN\left( hh\to t^-\bar{t}^+\right) &=  \mN\left(hh\to t^+\bar{t}^-\right) = 0, 
\end{align}
where%
\begin{multline}
  \mN \left( hh\to t^+ \bar{t}^+\right) = -2\sqrt{3}c_{2}\frac{M_t\sqrt{s}}{v^2}   \\
 + \sqrt{3}\frac{s}{v^{2}}\left[ \frac{g'^r_t(\mu)}{2}-\frac{3C'_t}{32\pi^2}%
\left( 2-\log\frac{-s}{\mu^2}\right) \right]  \frac{M_t\sqrt{s}}{v^2}.  \label{AmplitudeHiggsII}
\end{multline}

\section{Helicity amplitudes}
\label{sec:helicity}

In order to study the unitarity of the strongly interacting $\omega\omega$, $hh$ and  $t\bar{t}$ processes it is quite convenient to consider partial waves of the corresponding helicity amplitudes, as the unitarity relations do not couple different $J$ nor custodial isospin $I$. For example, for elastic Goldstone-boson scattering $\omega\omega\to \omega\omega$ there are three custodial isospin $A_I$ amplitudes $\left(I=0,1,2\right)$, analogous to those in pion-pion scattering in hadron physics,
\begin{align}
A_0\left( s,t,u\right) &= 3A\left( s,t,u\right) +A\left(t,s,u\right) +A\left( u,t,s\right) \nonumber\\
A_1\left( s,t,u\right) &=  A\left( t,s,u\right) -A\left(u,t,s\right) \nonumber\\
A_2\left( s,t,u\right) &=  A\left( t,s,u\right) +A\left(u,t,s\right) ,
\end{align}
which are defined in terms of the amplitude
\begin{multline}
  \mA (\omega_i\omega_j\to\omega_k\omega_l)= \\
  A(s,t,u)\delta_{ij}\delta_{kl} + A(t,s,u)\delta_{ik}\delta_{jl} + A(u,t,s)\delta_{il}\delta_{jk}.
\end{multline}
These amplitudes can be expanded as
\begin{equation}
A=A^{(0)} + A^{(1)} + \dots = A^{(0)}+A_{\rm tree}^{(1)} + A_{\rm loop}^{(1)} +\dots
\end{equation}
The projection over definite orbital angular momentum (the GBs carry zero spin) is then 
\begin{equation}
A_{IJ}^{(0)}\left( s\right) =\frac{1}{64\pi }%
\int_{-1}^{1}d\left( \cos \theta \right) P_{J}\left( \cos \theta \right)
A_{I}\left( s,t,u\right) .
\end{equation}
These partial waves also accept a chiral expansion
\begin{equation}
A_{IJ}=A_{IJ}^{(0)}+A_{IJ}^{(1)}+\dots,
\end{equation}
which takes the general form%
\begin{align}
A_{IJ}^{(0)}\left(s\right) &= Ks \\
A_{IJ}^{(1)}\left(s\right) &= \left(B(\mu) +D\log\frac{s}{\mu^2}+E\log\frac{-s}{\mu^2}\right) s^2 .
\end{align}

The constants $K$, $D$ and $E$ and the function $B(\mu)$ depend on the different channels $IJ=00;11;20;02;22$, as is shown in~\cite{Delgado:2013hxa,Delgado:2015kxa}. We will use the notation of that paper for the inelastic and pure-$h$ scattering reactions too. As $A_{IJ}\left(s\right)$ must be scale independent we have
\begin{equation}
    B(\mu) =B(\mu_0) + (D+E) \log\left(\frac{\mu^2}{\mu_0^2}\right) .
\end{equation}

This $B(\mu )$ function depends on the NLO chiral constants ($a$, $b$, $a_4$, $a_5$, etc.) and from now on we omit the superindices $r$ on the renormalized coupling constants for simplicity.

Since the Higgs boson is assigned zero custodial isospin, $\omega\omega\to hh$ and $hh\to hh$ occur only in the isospin zero channel $I=0$. The corresponding  partial waves can also be expanded as
\begin{equation}
    M_{J}=M_{J}^{(0)}+M_J^{(1)}+\dots
\end{equation}
and
\begin{equation}
    T_{J}=T_{J}^{(0)}+T_{J}^{(1)}+\dots,
\end{equation}
respectively. 
Both $\omega\omega$ and $hh$ may couple to the $t\bar{t}$ state. 
The $hh$ pair is always produced in an $I=0$ state as $h$ is a custodial symmetry singlet. On the other hand, as $t$ is a member of a custodial isospin doublet $(t,b)^T$, a $t\bar{t}$ doublet can be projected to both $I=0$ and $I=1$, with
\begin{equation}
    \ket{I=0, I_z=0} = \frac{1}{\sqrt{2}}(\ket{t\bar t} + \ket{b\bar b})
\end{equation}
and
\begin{equation}
    \ket{I=1, I_z=0} = \frac{1}{\sqrt{2}}(\ket{t\bar t} - \ket{b\bar b})\ .
\end{equation}
The $\omega\omega$ state with $I=0$ is defined as
\begin{equation}
    \ket{I=0} = \frac{1}{\sqrt{3}}\sum_i \ket{\omega^i\omega^i}\ .
\end{equation}

As the $t$ and $b$ quarks interact proportionally to their masses which are so different, the reactions involving fermions considered here are not custodial invariant (and in fact we are neglecting all the time $b \bar b$ pair production since $M_t \gg M_b \simeq 0$). The initial $I=0$ $\omega\omega$ or $hh$ states couple to $\ket{t\bar t}$ in a superposition of $\ket{I=0, I_z=0}$ and $\ket{I=1, I_z=0}$. 

If we concentrate in the $J=0$ case, the parity of the $I=J=0$ $\omega\omega$ or $hh$ pairs is positive; so must be that of the $t\bar{t}$ state. The reason is that, though Eq.~(\ref{EFTLagrangianexpanded}) contains parity-violating terms, the Feynman diagrams in Fig.~\ref{fig:OneWW} and \ref{fig:Higgoneloop2} only employ the parity-conserving pieces of that equation.

\begin{figure}[h]\centering
\tikzsetnextfilename{chiral}
\begin{tikzpicture}[
        scale=1.,
        axis/.style={very thick, ->, >=stealth', line join=miter},
        dot_red/.style={circle,fill=red,minimum size=4pt,inner sep=0pt,
            outer sep=-1pt},
        dot_blue/.style={circle,fill=blue,minimum size=4pt,inner sep=0pt,
            outer sep=-1pt},
    ]

\draw[axis,<->] (5,0) node(xline)[right] {$M_t$} -| (0,5) node(yline)[above] {$s$};   

\draw (1, 0.1) -- (1, -0.1) node [below] {$M_t\phantom{^1}$};
\foreach \x in {2,...,4}
    \draw (\x, 0.1) -- (\x, -0.1) node [below] {$M_t^\x$};

\draw (0.1, 1) -- (-0.1, 1) node [left] {$s^{1/2}$};
\draw (0.1, 2) -- (-0.1, 2) node [left] {$s\phantom{^{2/2}}$};   
\draw (0.1, 3) -- (-0.1, 3) node [left] {$s^{3/2}$};
\draw (0.1, 4) -- (-0.1, 4) node [left] {$s^{2\phantom{/2}}$};

\draw (-0.1,-0.1) node[below left] {$0$};

\node[dot_red,label=above right:LO] at (0,2) {};
\node[dot_red,label=above right:NLO] at (0,4) {};

\node[dot_blue,label=above right:LO] at (1,1) {};
\node[dot_blue,label=above right:NLO] at (1,3) {};

\node[dot_red,label=right:EWSBS alone] at (4,4) {};
\node[dot_blue,label=right:EWSBS+$t\bar{t}$] at (4,3) {};

\end{tikzpicture}
\caption{Chiral $(M_t,\sqrt{s})$ counting. Note that the HEFT is perturbative in $M_t$, but requires unitarization in $s$ to reach the resonance region.\label{fig:counting}}
\end{figure}
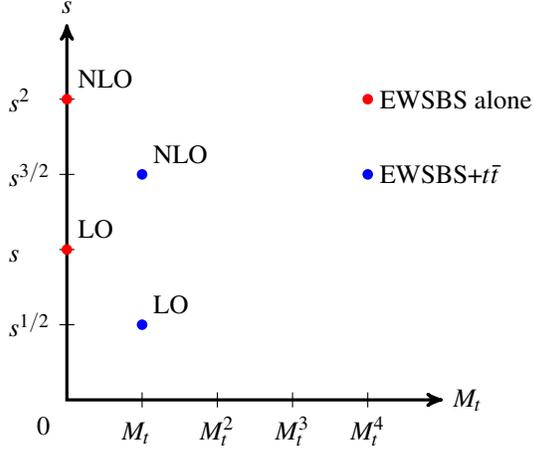

As the product of intrinsic fermion and antifermion parities is $P=-1$, their relative  orbital angular momentum must be $L=1$ (to obtain $P=+1$) and $J=0$ then needs spin $S=1$. Therefore we denote the $t\bar t$ helicity states by $\ket{\lambda_1,\lambda_2}$ and build up the $S=1$, $S_z=0$ state
\begin{equation} \label{spin1ttbar}
   \ket{S=1, S_z=0} = \frac{1}{\sqrt{2}}(\ket{+,+} - \ket{-,-}).
\end{equation}
The orthogonal spin state $\ket{S=0, S_z=0}$, having negative parity in an s-wave, does not couple to the $I=0$ $\omega\omega$ or $hh$ states.
Putting it in a p-wave would entail one more order in the chiral $(M_t,\sqrt{s})$ counting (depicted in Fig.~\ref{fig:counting}).

\section{Partial waves in perturbation theory}
\label{sec:partialwaves}

For the $I=0$ projection of the processes computed in this work we introduce the partial wave helicity amplitudes
\begin{align}
 Q^{J}_{\lambda_{1}\lambda_{2}}(s)=\frac{1}{64\pi^{2}}\int d\Omega \mD_{0\lambda}^{J}\left(\phi,\theta,-\phi\right)
 \mQ\left( \omega\omega\to t^{\lambda_{1}}\bar{t}^{\lambda_{2}}\right)
\end{align}   
and
\begin{align}
 N^{J}_{\lambda_{1}\lambda_{2}}(s)=\frac{1}{64\pi^{2}}\int d\Omega \mD_{0\lambda}^{J}\left(\phi,\theta,-\phi\right)
 \mN\left( h h \to t^{\lambda_{1}}\bar{t}^{\lambda_{2}}\right),
\end{align}   
where we only need the case $J=0$ and $\lambda=\lambda_1-\lambda_2=0$ (at an energy lower than intrinsic new physics scales in the 1 TeV region, only one or at most a few partial waves suffice to accurately represent the whole amplitude). Then, the rotation is simply represented  by the identity matrix. These partial waves are trivially obtained from our amplitudes in Eqs.~(\ref{AmplitudeGoldstone}-\ref{AmplitudeGoldstone3}) and~(\ref{AmplitudeHiggsIIpre}-\ref{AmplitudeHiggsII}). Taking into account Eq.~(\ref{AmplitudeGoldstone2}),
\begin{equation}
   Q^0_{++}=-Q^0_{--}.
\end{equation} 
Then, the partial wave corresponding to the $\ket{S=1,S_z=0}$ $t\bar t$ state in Eq.~(\ref{spin1ttbar}) is given by
\begin{equation}
   Q = \frac{1}{\sqrt{2}}(Q^0_{++}-Q^0_{--})=\sqrt{2}Q^0_{++}.
\end{equation} 
This partial wave can be expanded as
\begin{equation}
   Q=Q^{(0)}+Q^{(1)}+\dots,
\end{equation}
where the first two contributions to $Q$ have the form
\begin{align}
Q^{(0)}\left( s\right) &= K^Q\sqrt{s}M_t, \label{CoeffI} \\
Q^{(1)}\left( s\right) &= \left( B^Q\left( \mu \right) + E^Q\log\frac{-s}{\mu^2}\right) s\sqrt{s}M_t, \label{CoeffII}
\end{align}
respectively. Projecting Eq.~(\ref{Qdecomposed}), the coefficients are given by
\begin{align}
K^Q                 &=  \frac{3}{16\pi v^2}\left( 1-ac_1\right) , \\
B^Q\left(\mu\right) &=  \frac{3}{16\pi v^4}\left[\frac{g_t(\mu)}{2}+\frac{C_t}{8\pi^2}\right] , \\
E^Q                 &= -\frac{3}{16\pi v^4}\frac{C_t}{%
16\pi ^{2}}.
\end{align}

\begin{figure}
    \centering
    \includegraphics[width=0.95\columnwidth]{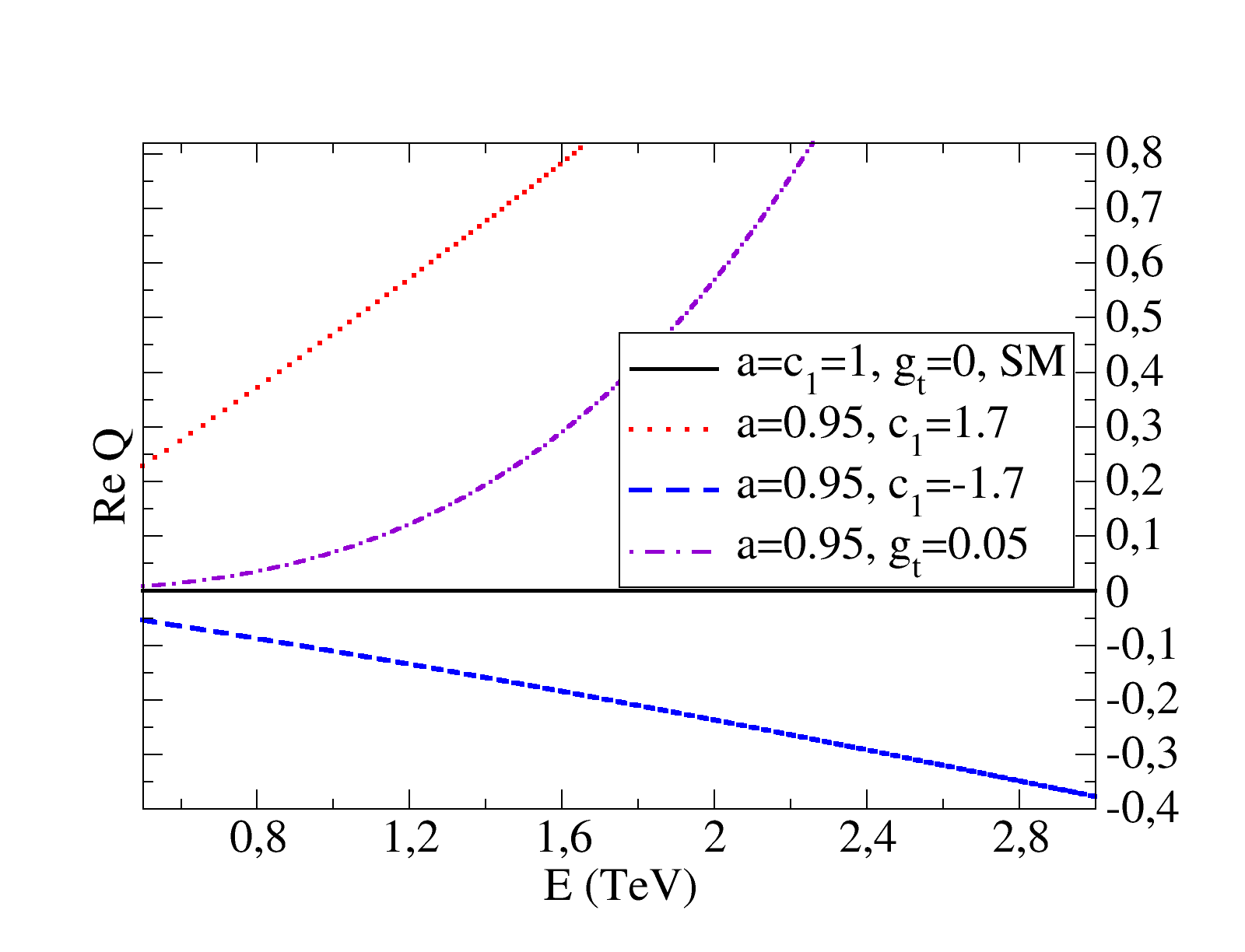}
    \caption{Real part of the $Q (\omega\omega\to t\bar{t})$ $I=J=0$ amplitude for $b=a^2=1$ as in the Standard Model and for departures thereof driven by $c_1$ (linear in $\sqrt{s}$) and $g_t$, with a slightly different value of $a$ but maintaining $b=a^2$, and $\mu=3$ TeV. \label{fig:realpartsPertTh}}
\end{figure}

\begin{figure}
    \centering
    \includegraphics[width=0.95\columnwidth]{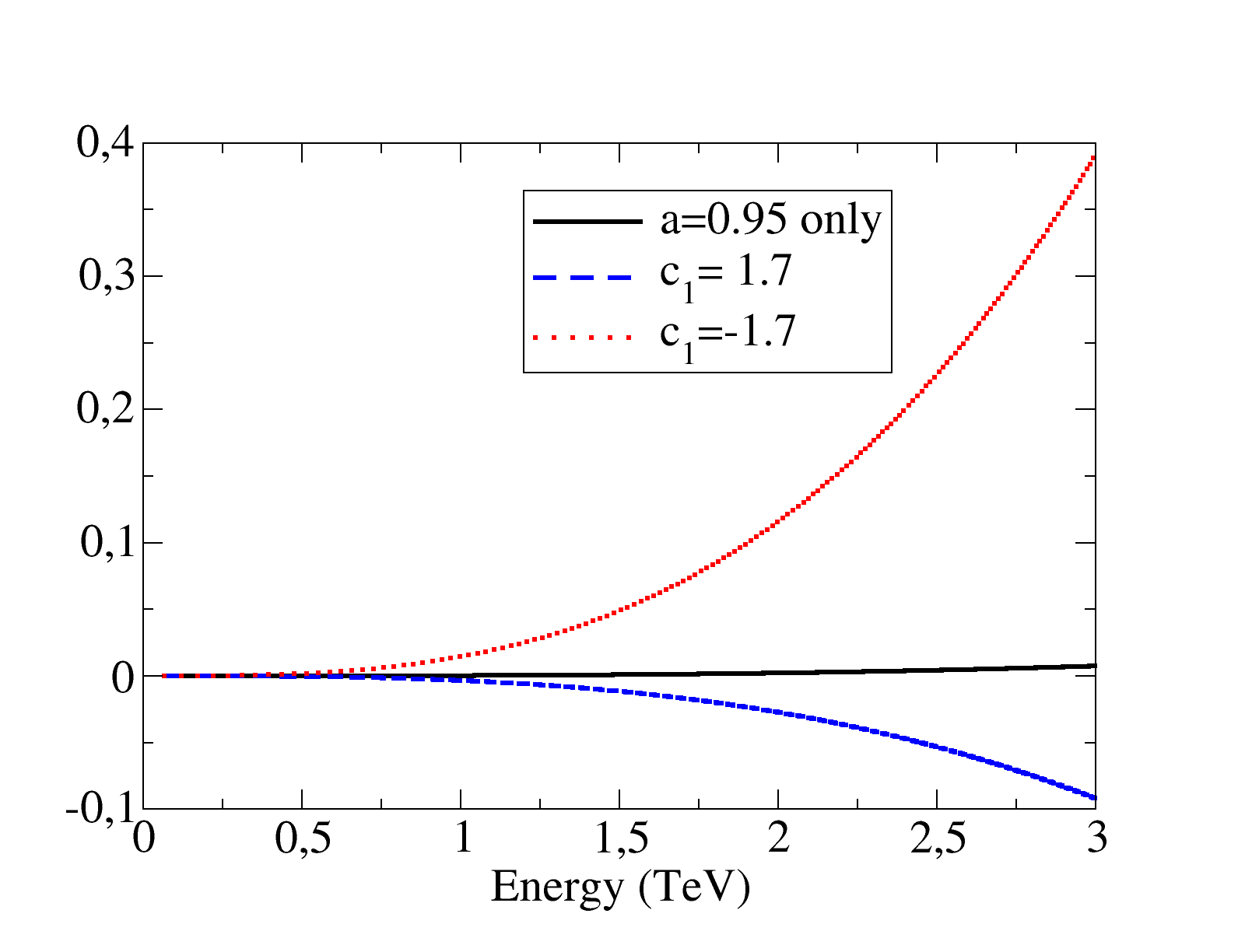}
    \caption{Imaginary part of the $Q (\omega\omega\to t\bar{t})$ $I=J=0$ amplitude for $b=a^2$ but $a=0.95$ slightly off the Standard Model, and for larger departures thereof driven by $c_1$ (linear in $\sqrt{s}$). \label{fig:imagpartsPertTh} }
\end{figure}

The real part of this scalar partial-wave is shown in Fig.~\ref{fig:realpartsPertTh}.
We can see in the figure the effect of the parameters $g_t$ and $c_1$. At present, $g_t$ is not very constrained, due to the low cross section of $hh$ production. But, according to~\cite{Buchalla:2015qju}, $c_1\in (1.0,\,1.7)$ at 2-$\sigma$ confidence level. For comparison, we also give the line corresponding to $c_1=1$, $g_t=0$ (at $\mu=3$TeV) and with $a=1$, so that the coupling to the $\omega\omega$ sector is as in the Standard Model. 
As visible, the amplitude may grow with $\sqrt{s}$ and may eventually violate perturbative unitarity (see section~\ref{sec:unitarization} below for an extensive discussion). The parameters from the top sector can easily enhance this behavior: for example, a value $g_t=0.03-0.05$ will already cause trouble with unitarity below $3\,{\rm TeV}$, as the amplitude is seen to approach 1 rapidly. Likewise, example imaginary parts of this partial wave are shown in Fig.~\ref{fig:imagpartsPertTh} (as this is not an elastic amplitude, the imaginary part can actually be negative depending on the parameter set).

In a similar way it is possible to obtain the $J=0$ partial wave for the  $hh\to t\bar{t}$ reaction, also accepting the chiral expansion
\begin{equation}
N=\sqrt{2}N^0_{++}=N^{(0)}+N^{(1)}+\dots,
\end{equation}
where the first two terms share the general form of Eq.~(\ref{CoeffI}-\ref{CoeffII}),
\begin{align}
N^{(0)}\left(s\right) &= K^N\sqrt{s}M_t, \label{ExpansionhhI} \\
N^{(1)}\left(s\right) &= \left( B^N(\mu) + E^N\log\frac{-s}{\mu^2}\right) s\sqrt{s}M_t,  \label{ExpansionhhII}
\end{align}
and the constants are given by
\begin{align}
K^N                 &= -\frac{\sqrt{3}c_2}{8\pi v^2}, \\
B^N\left(\mu\right) &=  \frac{\sqrt{3}}{16\pi v^4}\left(\frac{g'_t(\mu)}{2}-\frac{3C'_t}{16\pi^2}\right), \\
E^N                 &=  \frac{\sqrt{3}}{16\pi v^4}\frac{3C'_t}{32\pi^2}.
\end{align}

With this partial wave it is possible to describe $J=0$ scattering including the $\omega\omega$, $hh$ and $t \bar t$ states. We collect them all in a partial-wave amplitude-matrix, in the order just quoted,
\begin{equation}
F_{J=0} = \left( 
\begin{array}{ccc}
   A_{00} & M_{0} & Q \\ 
   M_{0} & T_{0} & N \\ 
   Q & N & S  \\ 
\end{array}%
\right) 
\end{equation}
where $S$ is the appropriate $t \bar t \to t \bar t $ partial wave. This partial wave is of order $M_t^2/v^2$, one  order higher than we have retained, so we may consistently set it to zero against others that are $M_t\sqrt{s}/(v^2)$.  As the interactions considered here are $T$-reversal invariant, this matrix is symmetric.   Each of the elements has a right unitary cut starting at $s=0$ associated with the threshold for producing  $\omega\omega$, $hh$ and $t \bar t$ (which are all considered massless here in accordance with the use of the Equivalence Theorem in the mid- to high-energy region). The physical partial waves have support on this cut along $s=E^2+i\epsilon$, where $E$ is the reaction's  center of mass energy. For these physical values, the unitarity condition for the $F$ matrix reads
\begin{equation} \label{matrixunitarity}
\Imag F=FF^\dagger .
\end{equation}
This matrix equation can be expanded to
\begin{subequations}\label{eq:unitar:general}
\begin{align}
   \Imag A &= \lvert A\rvert^2 + \lvert M\rvert^2 + \dots \label{eq:unitar:first}\\
   \Imag M &= A M^* + M T^* + \dots\\
   \Imag T &= \lvert M\rvert^2 + \lvert T\rvert^2 + \dots \\
   \Imag Q &= A Q^* + MN^* + \dots \label{chLagr:UnitProceduresWW:Qcoupled2:unitarQ}\\
   \Imag N &= MQ^* + TN^* + \dots \label{newunitarity2}\\
   \Imag S &= 0 + \dots \label{newunitarity3}
\end{align}
\end{subequations}
The ellipsis stand for third terms that are higher order in $M_t/v$. Thus, this system is only approximately equivalent to exact unitarity. As shown in the appendix (Fig.~\ref{fig:unitaritycoupledchannels}) our unitarization method in Eq.~(\ref{chLagr:UnitProceduresWW:Qcoupled2:unitar:raw}), as implemented on a computer, satisfies this system with very good accuracy.

However, the NLO computations shown in Refs.~\cite{Delgado:2013hxa,Delgado:2014dxa} for $A(s)$, $M(s)$ and $T(s)$; and those of this work for $Q(s)$ and $N(s)$, are unitary only in the perturbative sense. We expand the matrix $F$, using an obvious notation, as
\begin{equation}
   F=F^{(0)}+F^{(1)}+\dots
\end{equation} 
Substitution in Eq.~(\ref{matrixunitarity}) yields the perturbative unitarity relations
\begin{equation}
  \Imag F^{(1)}=F^{(0)}F^{(0)},
\end{equation}
The set of Eq.~(\ref{eq:unitar:first}-\ref{newunitarity3}) become (noting that the second, third and fourth equations simplify, as the last term drops because $T^{(0)}_0=0$ for the scalar $hh\to hh$ channel),
\begin{subequations}
\begin{align}
\Imag A_{00}^{(1)} &= \left\lvert A_{00}^{(0)}\right\rvert^2 + \left\lvert M_{0}^{(0)}\right\rvert^2 \label{pertunit1}\\
\Imag M_{0}^{(1)}  &= A_{00}^{(0)} M_{0}^{(0)} \label{pertunit2}\\
\Imag T_{0}^{(1)}  &=  \left\lvert M_{0}^{(0)} \right\rvert^2 \label{pertunit3}\\
\Imag Q^{(1)}      &= A_{00}^{(0)} Q^{(0)}  + M^{(0)}_0 N^{(0)}  \label{pertunit4}\\
\Imag N^{(1)}      &= M_{0}^{(0)} Q^{(0)} \label{pertunit5}\\
\Imag S^{(0)}      &= 0  . \label{pertunit6}
\end{align}
\end{subequations}
The first three equations were obtained in \cite{Delgado:2013hxa,Delgado:2014dxa}. The last two are equivalent to
\begin{align}
 -\pi E^Q &= K K^Q + K_0' K^N, \nonumber \\
 -\pi E^N &=   K_0'  K^Q
\end{align}
(with $K'_0$ the LO constant in the $M$ amplitude).

\begin{figure}
    \centering
    \includegraphics[width=0.95\columnwidth]{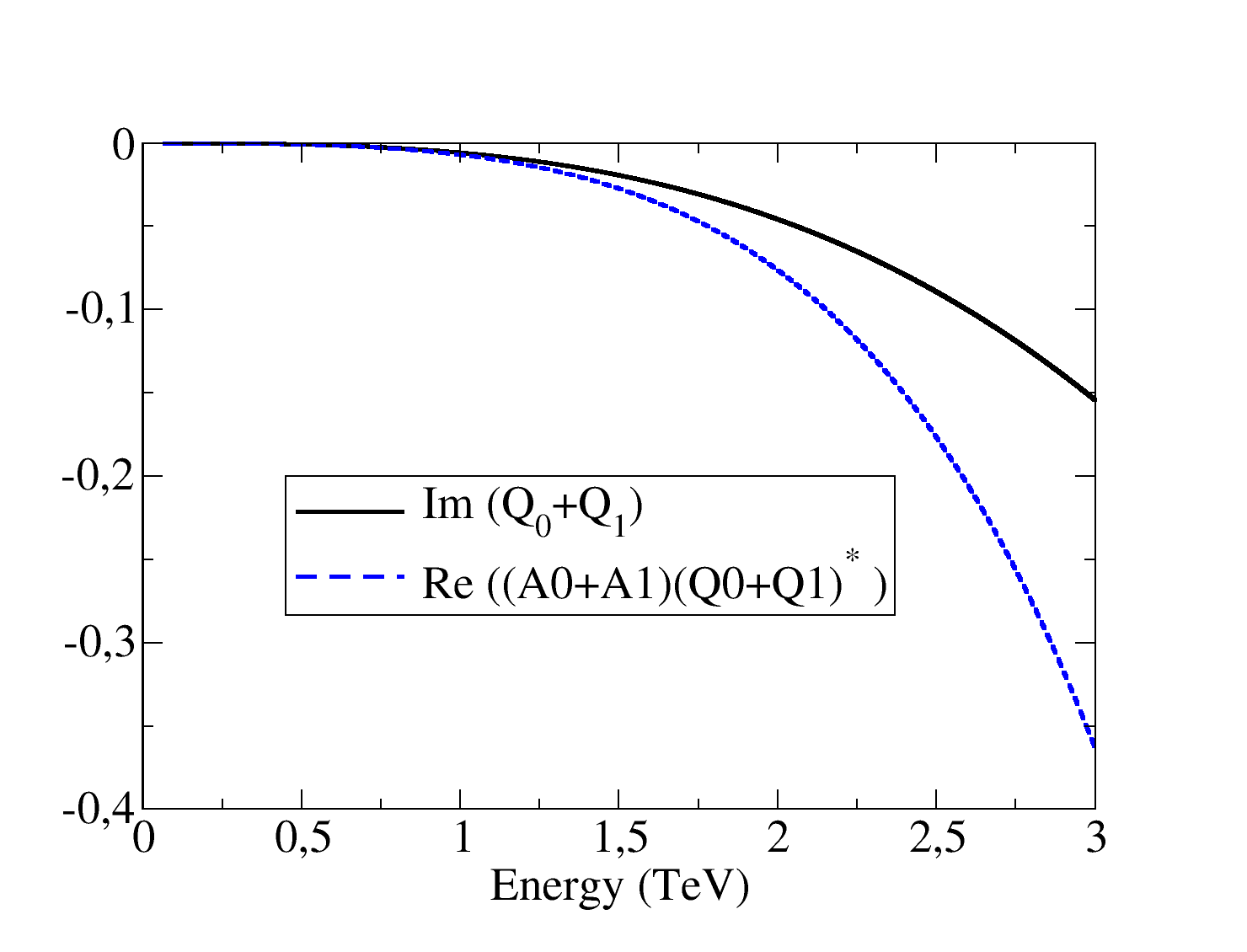}
    \caption{Perturbative unitarity: the imaginary part of the $\omega\omega\to t\bar{t}$ amplitude Q
    does not equal $AQ^*$ as it should if both are taken to NLO, satisfying instead Eq.~(\ref{pertunit4}). Parameters: $a=0.9$, $b=a^2$, $c_1=1.7$, $g_t(\mu=3\,{\rm TeV})=0$. \label{fig:unitaritycheck1}}
\end{figure}

These two equations can be explicitly checked and are a very non trivial consistency test of the results found in this work. However, as Fig.~\ref{fig:unitaritycheck1} shows, perturbative unitarity in a chiral expansion separates from exact unitarity as energy increases, invalidating perturbation theory.


\section{Partial-wave unitarization}%
\label{sec:unitarization}
In this section we will show how it is possible to promote the perturbative unitarity obtained from the one-loop computations to the exact (in a sense that will be clarified below) unitarity. To show how that can be done we will start first with the particular case $a^2=b$. From the results in  \cite{ Delgado:2013hxa,Delgado:2014dxa} it is very easy to realize that the $\omega\omega\to hh$ amplitude vanishes, and $hh$ decouples from $\omega\omega$ elastic scattering. Thus, the reaction matrix can be written in $2\times 2$ form as
\begin{equation}
   F = %
     \begin{pmatrix}
        A_{00} & Q_0 \\
        Q_0 & S_0
     \end{pmatrix}
     \equiv %
     \begin{pmatrix}
        A & Q \\
        Q & S
     \end{pmatrix},
\end{equation}
where $A_{00}$, $Q_0$ and $S_0$ are the elastic $\omega\omega\to\omega\omega$, cross-channel $\omega\omega\to t\bar{t}$, and elastic $t\bar{t}\to t\bar{t}$ $I=J=0$ partial waves respectively. As we saw in the previous section the $A$, $Q$ and $S$ amplitudes can be expanded as
\begin{subequations}
\begin{align}\label{chLagr:UnitProceduresWW:Qcoupled1}
   A &= A^{(0)} + A^{(1)} + ... & &\sim 1  \\
   Q &= Q^{(0)} + Q^{(1)} + ... & &\sim \mO\left(\frac{M_t}{v}\right) \label{chLagr:UnitProceduresWW:Qcoupled1:Q}\\
   S &= S^{(0)} + ...           & &\sim \mO\left(\frac{M_t^2}{v^2}\right) .
\end{align}
\end{subequations}
On the (RC) right cut (the physical region) the unitarity relation $\Imag F = F F^\dagger$  applies, entailing
\begin{subequations}
\begin{align}
   \Imag A &= \lvert  A \rvert^2 + \mO\left(\frac{M_t^2}{v^2}\right) \label{elasticunit}\\
   \Imag Q &= A Q^* + \mO\left(\frac{M_t^3}{v^3}\right) \label{chLagr:UnitProceduresWW:Qcoupled1:unitarQ}\\
   \Imag S &= 0 + \mO\left(\frac{M_t^2}{v^2}\right) .
\end{align}
\end{subequations}
These equations, if expanded in perturbation theory, return Eqs.~(\ref{pertunit1}), (\ref{pertunit4}) and (\ref{pertunit6}) (setting $M=0$ there).
In order to fulfill these relations we proceed as follows. First, we consider the elastic scattering $\omega\omega$ amplitude $A$. As shown in~\cite{Delgado:2013hxa,Delgado:2014dxa}, Eq.~(\ref{pertunit4}) can be satisfied by using the Inverse Amplitude Method (IAM) which introduces the unitarized amplitude
\begin{equation}\label{onechannelIAM}
\tilde{A}=\frac{(A^{(0)})^2}{A^{(0)}-A^{(1)}}. 
\end{equation}
This ensures elastic unitarity, $\Imag\tilde{A} = \tilde{A}\tilde{A}^*$, provided we have perturbative unitarity, i.e. $\Imag A^{(1)} = (A^{(0)})^2$, as is the case here. Now, in order to unitarize $Q$ we introduce
\begin{equation}\label{chLagr:UnitProceduresWW:Qcoupled1:unitar:raw}
   \tilde{Q} = Q^{(0)} + Q^{(1)}\frac{\tilde{A}}{A^{(0)}} .
\end{equation}
Again, it is very easy to show that this partial wave fulfills the unitarity relation $\Imag \tilde Q = A Q^*$ by using the perturbative result $\Imag Q^{(1)}=Q^{(0)}A^{(0)}$ as follows,
\begin{align}
  \left.\tilde{Q}\right\rvert_{\rm RC}   &=  Q^{(0)} + Q^{(1)}\frac{\tilde{A}}{A^{(0)}}   \nonumber\\
          &=  \left[Q^{(0)}\left(1-\frac{A^{(1)}}{A^{(0)}}\right) + Q^{(1)}\right]\frac{A^{(0)}}{A^{(0)}-A^{(1)}} \nonumber\\
          &=  \left[Q^{(0)} - \frac{Q^{(0)}}{A^{(0)}}\Real A^{(1)} + \Real Q^{(1)} \right]\frac{\tilde{A}}{A^{(0)}}
\end{align}
Thus, we have that
\begin{multline}
  \left.\Imag\tilde{Q}\right\rvert_{\rm RC} = %
    \left[Q^{(0)} - \frac{Q^{(0)}}{A^{(0)}}\Real A^{(1)} + \Real Q^{(1)} \right]\frac{\Imag\tilde{A}}{A^{(0)}} \\= %
    \left[Q^{(0)} - \frac{Q^{(0)}}{A^{(0)}}\Real A^{(1)} + \Real Q^{(1)} \right]\frac{\tilde{A}\tilde{A}^*}{A^{(0)}} = %
      \tilde{Q}\tilde{A}^*.
\end{multline}
Hence, we recover Eq.~(\ref{chLagr:UnitProceduresWW:Qcoupled1:unitarQ}) as announced. Therefore we are left with two unitarized amplitudes $\tilde A$ and $\tilde Q$ for the processes $\omega\omega\to\omega\omega$ and $\omega\omega\to t\bar t$, respectively. These amplitudes also respect the perturbative expansion
\begin{align}
 \tilde A &= A^{(0)} + A^{(1)} + \dots \nonumber\\
 \tilde Q &= Q^{(0)} + Q^{(1)} + \dots
\end{align}
and, in addition they feature the proper analytical structure on the whole complex plane. In particular they have a right (unitarity) cut and also the expected left cut. Moreover, they can be analytically extended to the second Riemann sheet beyond the unitarity cut and they can have poles there that can be understood as resonances (whether ``dynamic'' or ``intrinsic'')  developing in some regions of the chiral coupling space. Those resonances are typical of strongly interacting scenarios for the symmetry breaking sector of the SM and are under active research at the LHC. 

The unitarity condition for the $Q$ amplitude linking $\omega\omega$ and $t\bar{t}$
introduced in Eq.~(\ref{chLagr:UnitProceduresWW:Qcoupled1:unitar:raw})
can now be checked numerically, and we have do so (not shown).
Our numeric precision is, for the entire energy interval of interest up to $3\,{\rm TeV}$, of order $10^{-5}$ without any particular effort (and this small error probably stems from our setting $b$ not quite equal to $a^2$ to avoid numerical problems elsewhere, so that a tiny leak to the $hh$ channel may be there), 
so that Watson's final state theorem is well satisfied and the phase of the $Q$ amplitude is correctly set to that of the strongly interacting $A(\omega\omega\to\omega\omega)$.

\begin{figure} 
\centering
    \includegraphics[width=0.95\columnwidth]{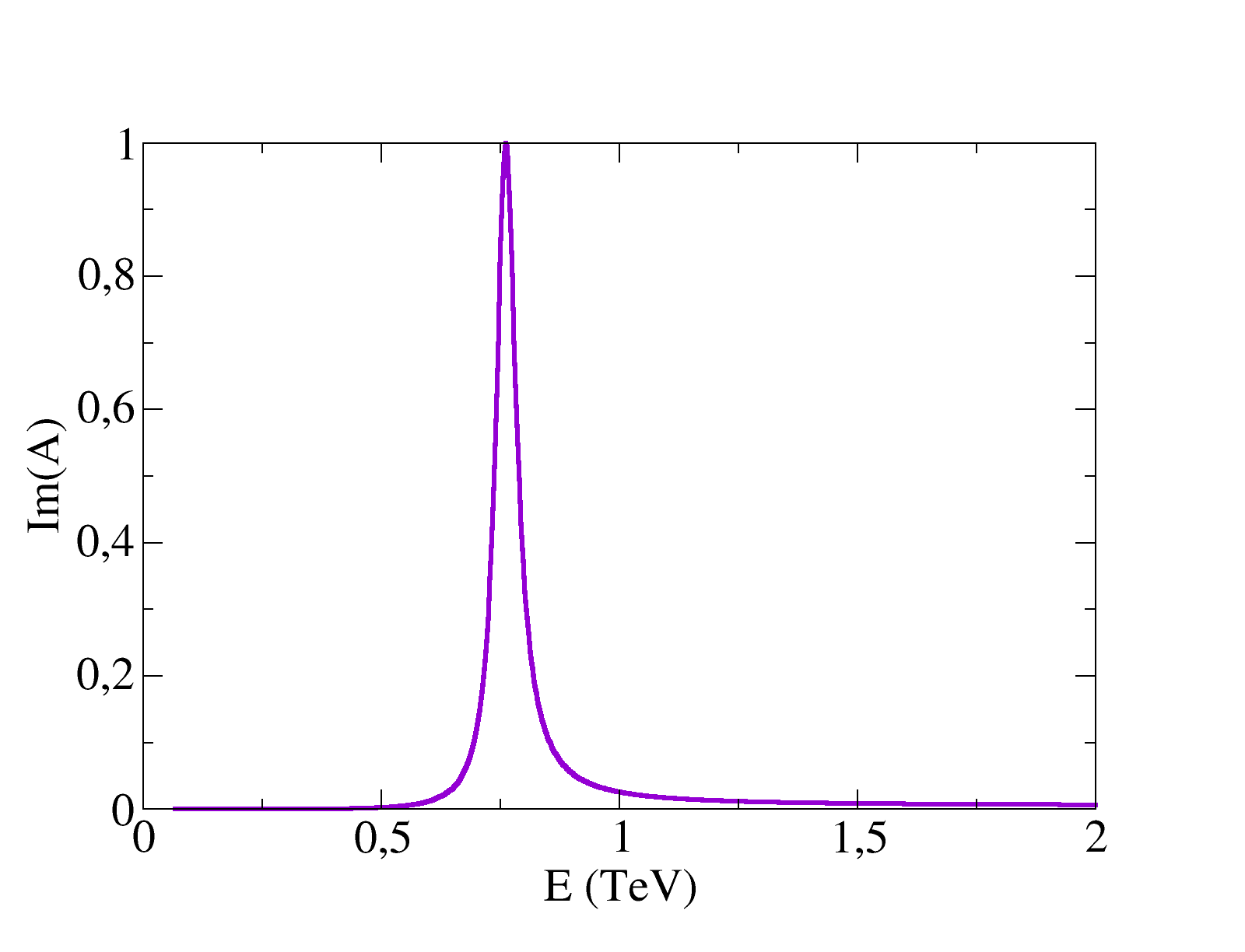}\\
    \includegraphics[width=0.95\columnwidth]{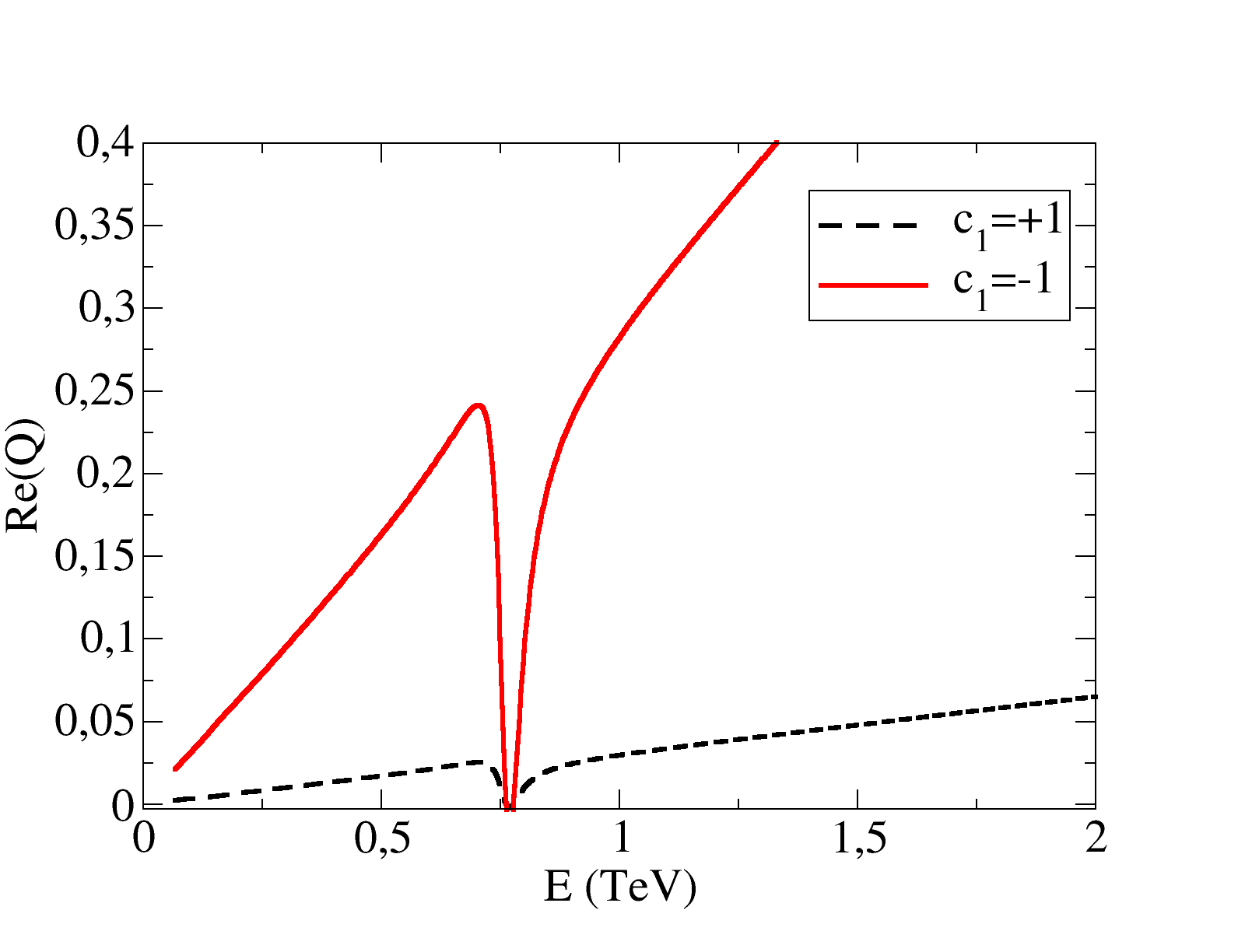}
    \caption{\label{fig:narrowelasticres}Top plot: a relatively narrow, scalar EWSBS resonance in the elastic $A$ ($\omega\omega\to\omega\omega$ amplitude). Lower plot: its appearance, for the indicated $c_1$, in the $t\bar{t}$ channel. Other parameters are indicated in the text. Note that this resonance interferes destructively with the background $Q$ amplitude.}
\end{figure}

We now exemplify the power of a method by generating a resonance in the elastic $A(\omega\omega\to\omega\omega)$ amplitude and feeding it to the $t\bar t$ channel. 
So that a comparison with many other theory works can be made, that were inspired by
a presumed narrow LHC excess at around 0.75 TeV, we choose $\mu\simeq M=750$GeV, $a=0.81$, $a_5=0.0023$, and all other parameters from the $\omega\omega$ sector as in the SM (particularly $a_4=0$ and $b\simeq a^2$). This generates a relatively narrow resonance with mass around 750 GeV, $\Gamma/M\simeq 0.06$ similar to what the community was considering before it was clear 
that it had been a statistical fluctuation.
The resonance is shown in the top plot of Fig.~\ref{fig:narrowelasticres}. The lower plot shows its effect on the real part of the $Q$ amplitude, where we have set differently from zero the parameter $c_1$ (to $\pm 1$).
As can be seen in the figure, the resonance, a textbook Breit-Wigner in the $A$ elastic channel (EWSBS) appears as a dip due to its interference with the background in the $Q(\omega\omega\to t\bar{t})$ amplitude. Of course, such dips will appear broadened and lessened after convolution with the parton distribution functions producing the top-antitop system, the hard kernels, and the reconstruction efficiency (and detector acceptance) of the final product decays. Though a full simulation is beyond the scope of this work, it is possible that they may be observable, providing a signal that is not so often expected (as practitionners often seek excess cross-sections).
A similar phenomenon has been observed by~\cite{Djouadi:2016ack} in interference between perturbative SM production $WW\to t\bar{t}$ (the equivalent of our $Q$ amplitude) and the 
s-channel production of $t\bar{t}$ via a new resonance of $\mO$(TeV) mass. The coincidence suggests that this may be a robust result. These interference phenomena of backgrounds and narrow resonances are well known in hadron physics~\cite{Harada:1995dc,Oller:1997ng} and it would be interesting to discover them in the EWSBS.

{
\newcommand{\QN}{\ensuremath{\begin{pmatrix} Q \\ N \end{pmatrix}}}
\newcommand{\tQN}{\ensuremath{\begin{pmatrix} \tilde{Q} \\ \tilde{N} \end{pmatrix}}}
\newcommand{\QNO}{\ensuremath{\begin{pmatrix} Q^{(0)} \\ N^{(0)} \end{pmatrix}}}
\newcommand{\QNI}{\ensuremath{\begin{pmatrix} Q^{(1)} \\ N^{(1)} \end{pmatrix}}}
The entire discussion can now be extended to the more general case  $a^2\neq b$ where the cross-channels $\omega\omega\to hh$ and $hh\to t\bar{t}$ are  active again. Then the reaction matrix is $3\times 3$,
\begin{equation}
   F = \begin{pmatrix}
          A_{00} & M_0 & Q \\
          M_0    & T_0 & N \\
          Q      & N   & S   
       \end{pmatrix} \equiv %
       \begin{pmatrix}
          A & M & Q \\
          M & T & N \\
          Q & N & S
       \end{pmatrix} .
\end{equation}
Here, $A_{00}$, $Q_0$ and $S_0$ are the scalar partial waves as in Eq.~(\ref{chLagr:UnitProceduresWW:Qcoupled1}) and following. 
And now, since the crossed channel amplitude $\omega\omega\to hh$ is present, we also include 
its $J=0$ partial wave, and those for $hh\to hh$ and $hh\to t\bar{t}$. All of them accept a chiral expansion as
\begin{align}
   F &= F^{(0)} + F^{(1)} + \dots, & \Imag F^{(0)}&\equiv 0 .
\end{align}
Once again, the partial waves $Q$, $N$ and $S$ are suppressed by $M_t/v$ factors. In particular,
\begin{align}\label{chLagr:UnitProceduresWW:Qcoupled2}
   X &= X^{(0)} + X^{(1)} + \dots & \sim & 1        \\
   Q &= Q^{(0)} + Q^{(1)} + \dots & \sim & \mO\left(\frac{M_t}{v}\right)  \\
   N &= N^{(0)} + N^{(1)} + \dots & \sim & \mO\left(\frac{M_t}{v}\right)  \\
   S &= S^{(0)} + \dots           & \sim & \mO\left(\frac{M_t^2}{v^2}\right)  ,
\end{align}
where $X= A$, $M$ or $T$. On the RC, the unitarity relation $\Imag F = F F^\dagger$ applies, which leads to the set of Eqs.~(\ref{eq:unitar:general}), where we omit terms suppressed by higher powers of $M_t/v$ in such a way that all equations are correct up to $\mO(M_t^2/v^2)$. This is essential to be able to decouple the unitarization of the WBGBs sector (the $A$, $M$ and $T$ partial waves) from the $t\bar{t}$ amplitudes. For the unitarization of the WBGBs sector we can use again the (coupled) IAM method. For this, we first define the $2\times 2$ matrix
\begin{equation}
   K \equiv \begin{pmatrix}
           A & M \\
           M & T
       \end{pmatrix},
\end{equation}
which as usual admits a chiral expansion $K=K^{(0)}+K^{(1)}+\dots$ with $\Imag K^{(1)}= K^{(0)}K^{(0)}$ on the RC (perturbative unitarity). The corresponding unitarized matrix $\tilde K$ is provided by the IAM method (basically, a dispersive analysis for this matrix that employs the chiral expansion on the LC and everywhere for small $s$, and exact two-channel unitarity on the RC), that generalizes Eq.~(\ref{onechannelIAM})
\begin{equation}
\tilde{K}=K^{(0)}(K^{(0)}-K^{(1)})^{-1} K^{(0)} \ .
\end{equation}
By construction, the unitarity relation $\Imag\tilde{K} = \tilde{K}\tilde{K}^\dagger$ holds. Now, the remaining unitarity conditions on the RC, Eqs.~(\ref{chLagr:UnitProceduresWW:Qcoupled2:unitarQ}-\ref{newunitarity3}) can be written in a condensed way as
\begin{equation}\label{chLagr:UnitProceduresWW:Qcoupled2:tmp0}
  \Imag\QN = K \QN^*.
\end{equation}
This can also be expanded in perturbation theory,
\begin{equation}\label{chLagr:UnitProceduresWW:Qcoupled2:tmp1}
  \Imag \QNI = K^{(0)}\QNO .
\end{equation}
A solution to Eq.~(\ref{chLagr:UnitProceduresWW:Qcoupled2:tmp0})  can be written down in generalizing the simpler $a^2= b$ discussion. The unitarized amplitudes are then (a simple demonstration is relegated to the appendix)
\begin{equation}\label{chLagr:UnitProceduresWW:Qcoupled2:unitar:raw}
\tQN = \QNO + \tilde{K} K_0^{-1}\QNI .
\end{equation}
Notice that we are using the notation $K_0\equiv K^{(0)}$, and that Eq.~(\ref{chLagr:UnitProceduresWW:Qcoupled2:unitar:raw}) is a generalization of Eq.~(\ref{chLagr:UnitProceduresWW:Qcoupled1:unitar:raw}). 
In the particular case $a^2=b$ we have $M^{(0)}=\tilde M=T^{(0)}=\tilde T = 0$ and we then recover the previous definitions of the unitarized $\tilde A$ and $\tilde Q$.
In the general case, the amplitudes obtained from Eq.~(\ref{chLagr:UnitProceduresWW:Qcoupled2:unitar:raw}) feature all the good properties mentioned above as analyticity in the whole complex plane, left and right cuts, the possibility for developing poles in the second Riemann sheet, etc.

\section{Discussion} \label{sec:discussion}

The possibility of coupling of the top-antitop quark pair to the longitudinal gauge bosons has long been considered~\cite{Veltman:1990wi}. We have here carried out a study that, while keeping perturbation theory relatively simple (by neglecting masses and transverse gauge couplings as well as lighter quarks, that is, concentrating on the electroweak symmetry breaking sector where new strong interactions appear), proceeds beyond it by implementing unitarity in the spirit of the final state interaction theorem.

In this article we have adopted an Effective Field Theory approach extending the ECL to incorporate a light Higgs, namely the HEFT, and coupled the resulting system to the top-antitop sector at NLO in a double expansion in $M_t/v$ and $\sqrt{s}/v$. 

It is clear that for large enough values of the $c_1$, $g_t$ or other parameters of the top sector, the coupling can be so intense than the approximation that all the strong interactions are contained in the EWSBS may fail. In that case, expressions such as Eq.~(\ref{chLagr:UnitProceduresWW:Qcoupled2:unitarQ})
are no longer a reliable guide, and a full coupled channel unitarization must be attempted, which we refrain from at the present time.
Fig.~\ref{fig:Mttestpertexp} gives a feeling as to when this is expected to happen.
\begin{figure}
    \centering
    \includegraphics[width=0.95\columnwidth]{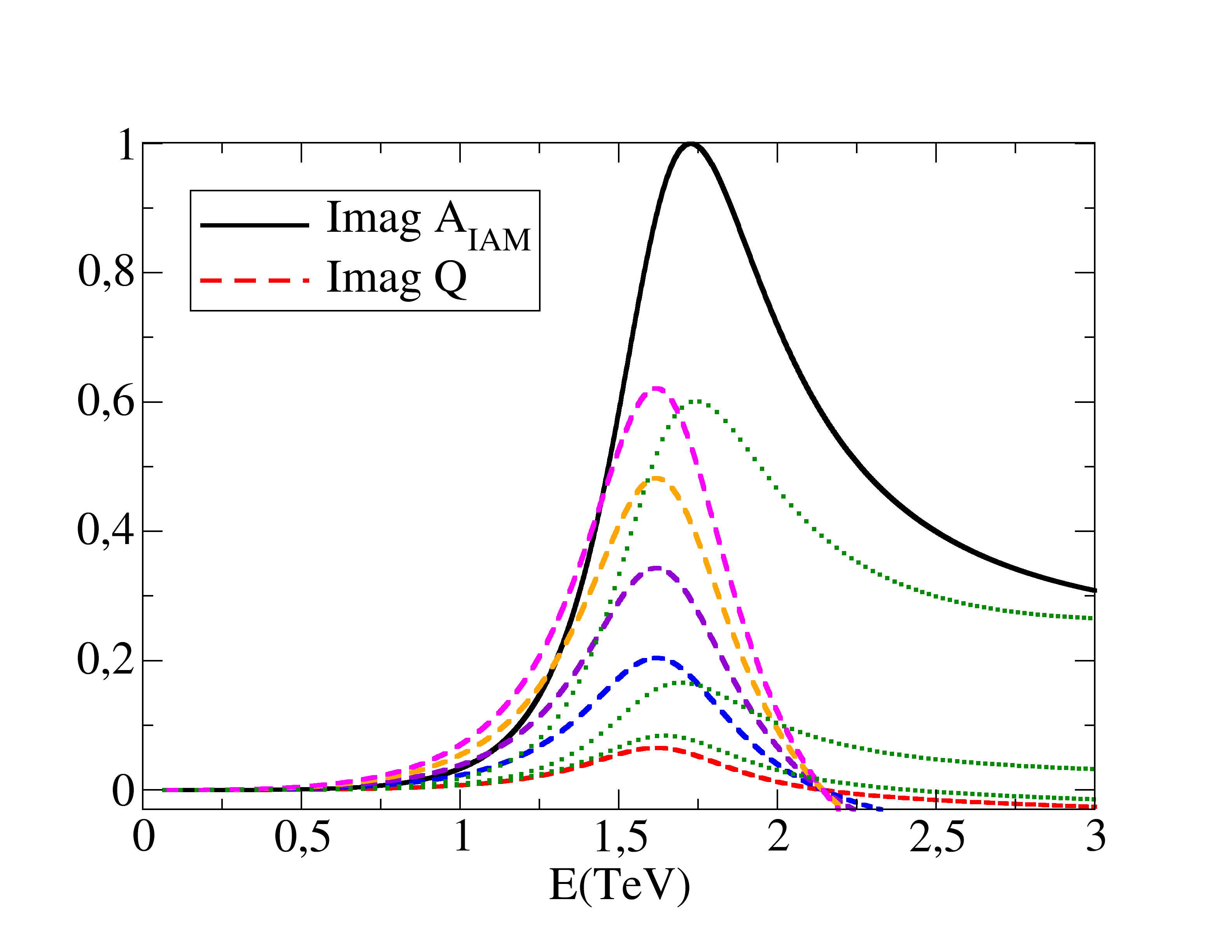}
    \caption{For high enough values of $c_1$ or $g_t$ (or other parameters not depicted), 
    the $M_t$ counting can be overruled. Solid line: $\Imag \tilde{A}_{IAM}(\omega\omega\to\omega\omega$) elastic amplitude. Others: $\Imag Q$. From top to bottom: dashed lines correspond to $c_1=1,0.5,0,-0.5,1$ respectively, while the dotted lines have been generated with $c_1=1$ but $g_t= 10^{-3},\ 5\times 10^{-3}$, and $2.5\times 10^{-2}$ respectively.
    In all cases the nonvanishing EWSBS parameters are $a=0.81$, $b=a^2$ as in the SM and $a_4(3\,{\rm TeV})=4\times 10^{-4}$.
    When $\Imag Q\sim\Imag\tilde{A}$, the $M_t/v$ expansion is no more a reliable guide and a full coupled-channel unitarization is necessary.}
    \label{fig:Mttestpertexp}
\end{figure} %
We see in the figure that for values of $c_1$ further than one and half units from its SM value ($c_1=1$) or  values of $g_t$ of order $0.01$, the interactions coupling $\omega\omega$ and $t\bar{t}$ become about 1/2 of the elastic $\omega\omega$ ones, and more care is required in studying the amplitudes.

As long as the top-sector parameters are smaller, we may use the natural counting represented above in Fig.~\ref{fig:counting}. 
Then, we have obtained the NLO amplitudes and studied their perturbative unitarity in Eqs.~(\ref{pertunit1}-\ref{pertunit6}) up to $M_t^2/v^2$ terms. The satisfaction of these unitarity relation deteriorates with increasing energy.

We have also shown the effect of employing those NLO scattering amplitudes from perturbation theory as the low-energy information for a dispersive analysis that can reach the resonance region ($E\sim 0.5-3$ TeV) as encoded in the Inverse Amplitude Method. 
This may prove useful if the LHC finds new resonances in the TeV region that it is exploring. It would be natural to take as starting point that any such resonances are related to electroweak symmetry breaking (otherwise why would they lie in this energy range), and their coupling to $t\bar{t}$ would be a promising alley of experimental investigation.

The IAM can reproduce broad, $\sigma$-like resonances as that depicted in Fig.~\ref{fig:Mttestpertexp}, driven by the LO parameter $a$, or narrow resonances such as that in Fig~\ref{fig:narrowelasticres}. We look forward to good-statistics LHC data to guide theory in the choice of HEFT parameters.

\section*{Acknowledgements}

The authors thank useful conversations with D.Espriu, M.J. Herrero and J.J.Sanz-Cillero, and the very constructive and useful comments of an anonymous referee. A.Dobado thanks the CERN TH-Unit for its hospitality during the time some important parts of this work were done. R.L. Delgado thanks the SLAC Theory Group for its hospitality and encouragement. The work has been supported by the Spanish grants No. UCM: 910309, MINECO:FPA2014-53375-C2-1-P and FPA2016-75654-C2-1-P, and by the grant MINECO:BES-2012- 056054 (R.L. Delgado).
A. Castillo is indebted to the \emph{Programa Nacional Doctoral of Colciencias-567} for its academic and financial support and also thanks kind hospitality and encouragement of the Group of Effective Theories in Modern Physics at the Universidad Complutense de Madrid. 
%

\section*{Appendix: unitarity in coupled channels}
Let us now prove that the amplitudes defined in Eq.~(\ref{chLagr:UnitProceduresWW:Qcoupled2:unitar:raw}) indeed satisfy the unitarity relations in Eq.~(\ref{chLagr:UnitProceduresWW:Qcoupled2:tmp0}) by taking the imaginary part of the former equation,
\begin{equation}\label{chLagr:UnitProceduresWW:Qcoupled2:tmp2}
   \Imag\tQN = \Imag\tilde{K}K_0^{-1}\Real\QNI + \Real\tilde{K} K_0^{-1}\Imag\QNI \ .
\end{equation}
Because of Eq.~(\ref{chLagr:UnitProceduresWW:Qcoupled2:tmp1}), Eq.~(\ref{chLagr:UnitProceduresWW:Qcoupled2:tmp2}) turns into
\begin{align}
 &  \Imag \tilde \QN = \Imag \tilde{K}K_0^{-1} Re\QNI +  Re\tilde{K}\QNO \nonumber\\%
 &= \tilde{K}\QNO - i \Imag\tilde{K}\QNO + \Imag\tilde{K}K_0^{-1} Re\QNI \nonumber\\%
 &= \tilde{K}\QNO + \Imag\tilde{K}K_0^{-1} Re\QNI     \nonumber\\%
 &\quad- \Imag\tilde{K}K_0^{-1}\Imag\QNI \nonumber\\ %
 &= \tilde{K}\QNO + \tilde{K}\tilde{K}^*K_0^{-1}\QNI^* = \tilde{K}\tQN^*,
\end{align}
so that we recover Eq.~(\ref{chLagr:UnitProceduresWW:Qcoupled2:tmp0}), as was to be demonstrated.

Notice also that the unitarization of Eq.~(\ref{chLagr:UnitProceduresWW:Qcoupled2:unitar:raw}) reproduces the correct low-energy behavior given by the chiral expansion,
\begin{multline}
   \tQN = \QNO + (K^{(0)}+K^{(1)}+\dots)K_0^{-1}\QNI \\ = \QNO + \QNI + \dots
\end{multline}

Finally, an explicit form can be given by expanding Eq.~(\ref{chLagr:UnitProceduresWW:Qcoupled2:unitar:raw}); it is immediate to find
\begin{align}
   \tilde{Q} ={}&  Q^{(0)} 
                 + Q^{(1)}\frac{\tilde{A}T^{(0)}-\tilde{M}M^{(0)}}{A^{(0)}T^{(0)}-(M^{(0)})^2}   \nonumber\\   
               & + N^{(1)}\frac{\tilde{M}A^{(0)}-\tilde{A}M^{(0)}}{A^{(0)}T^{(0)}-(M^{(0)})^2} , \nonumber\\   
   \tilde{N} ={}&  N^{(0)}
                 + Q^{(1)}\frac{\tilde{M}T^{(0)}-\tilde{T}M^{(0)}}{A^{(0)}T^{(0)}-(M^{(0)})^2}   \nonumber\\   
               & + N^{(1)}\frac{\tilde{T}A^{(0)}-\tilde{M}M^{(0)}}{A^{(0)}T^{(0)}-(M^{(0)})^2} ,
\end{align}
where the unitarized partial waves $\tilde{A}$, $\tilde{M}$ and $\tilde{T}$, which correspond to the WBGBs sector, are computed with the IAM procedure. 

In Fig.~\ref{fig:unitaritycoupledchannels} we show a check of our approximate coupled-channel unitarity relations from Eq.~(\ref{eq:unitar:general}) above. The imaginary parts of $Q$ and $N$ linking the $\omega\omega$ and $hh$ channels to $t\bar{t}$, respectively, are shown against the two separate terms on the RHS of those equations and against their sum, which perfectly match the imaginary part. 
\begin{figure}
    \centering
    \includegraphics[width=0.9\columnwidth]{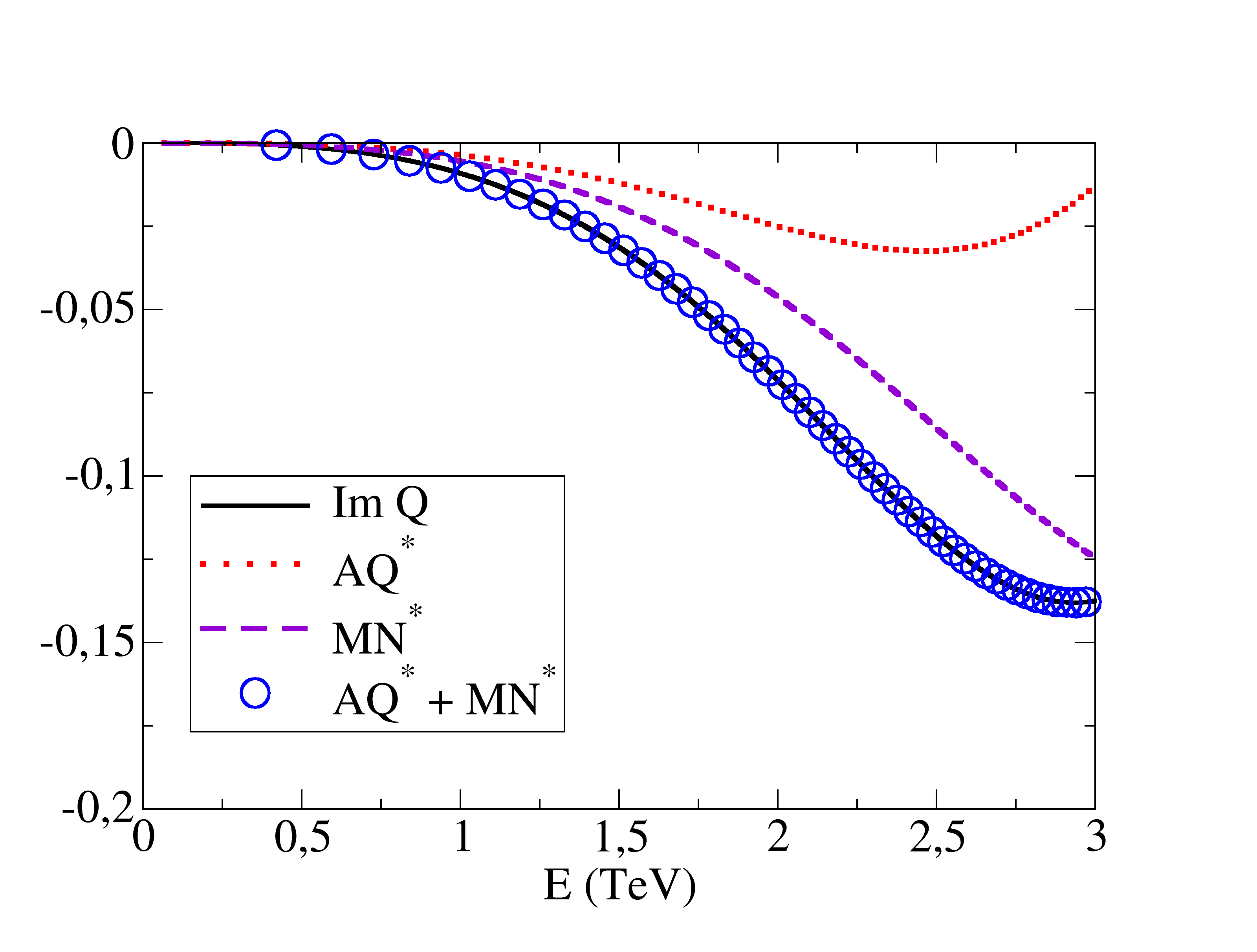}\\
    \includegraphics[width=0.9\columnwidth]{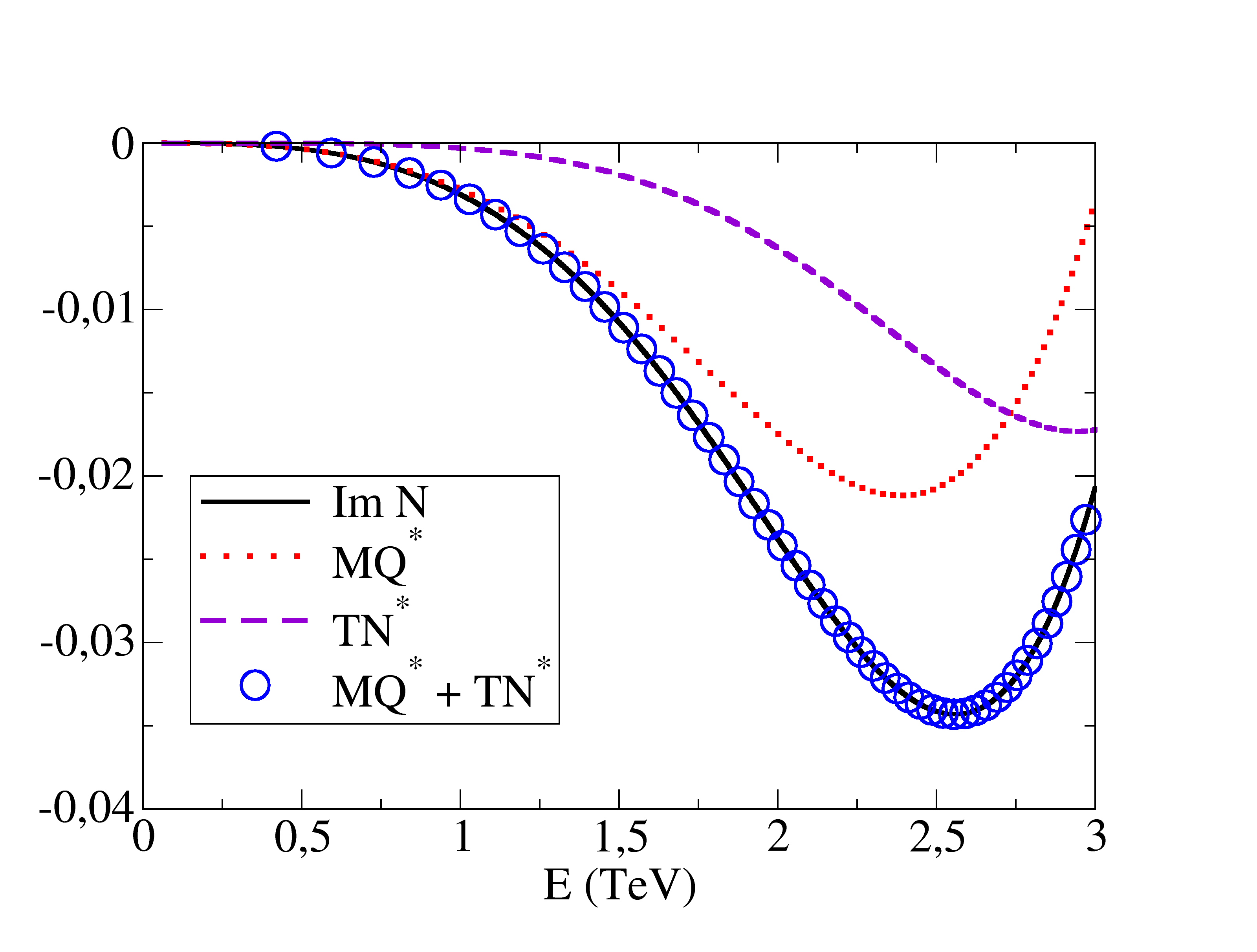}
    \caption{Tests of approximate coupled channel unitarity. The parameters used are $c_1 = 1.7$, $c_2 = 1$, $g_t=1$, $g_t'=0$ at $\mu=3\,{\rm TeV}$, with $a=0.95$ and $b=0.9a^2$.}
    \label{fig:unitaritycoupledchannels}
\end{figure}


\end{document}